\documentclass[aps,pre,twocolumn,showpacs]{revtex4}

\usepackage{graphicx, verbatim, amssymb, amsmath}
\usepackage{float}
\usepackage{subfig}
\usepackage{color}

\newcommand{\ket}[1]{\vert #1 \rangle} \newcommand{\bra}[1]{\langle #1 \vert}
\newcommand{\be}{\begin{equation}}
\newcommand{\ee}{\end{equation}}
\newcommand{\bae}{\begin{eqnarray}}
\newcommand{\eae}{\end{eqnarray}}
\newcommand{\average}[1]{\left \langle #1  \right\rangle}

\newcommand{\hli}{{\scriptscriptstyle HLI}}
\newcommand{\lli}{{\scriptscriptstyle LLI}}

\begin{document}

\title{Topology and energy transport in networks of interacting
photosynthetic complexes}

\author{Michele Allegra$^{1}$ $^{2}$, Paolo Giorda$^{1}$}

\affiliation{$^1$Institute for Scientific Interchange Foundation (ISI), Via Alassio 11/c, I-10126 Torino, Italy}

\affiliation{$^2$Dipartimento di Fisica dell'Universit\`a di Torino \& INFN, Sezione di Torino, via P. Giuria 1, I-10125 Torino, Italy}




\begin{abstract}
We address the role of topology in the energy transport process that occurs in networks of photosynthetic complexes.
We take inspiration from light harvesting networks present in purple bacteria and simulate an incoherent dissipative energy transport process on more general and abstract networks, considering both regular structures (Cayley trees and hyperbranched fractals) and randomly-generated ones. We focus on the the two primary light harvesting complexes of purple bacteria, i.e., the LH1 and LH2, and we use network-theoretical centrality measures in order to select different LH1 arrangements. We show that different choices cause significant differences in the transport efficiencies, and that for regular networks centrality measures allow to identify arrangements that ensure transport efficiencies which are better than those obtained with a random disposition of the complexes.
The optimal arrangements strongly depend on the dissipative nature of the dynamics and on the topological properties of the networks considered, and depending on the latter they are achieved by using global vs. local centrality measures. For randomly-generated networks a random arrangement of the complexes already provides efficient transport, and this suggests the process is strong with respect to limited amount of control in the structure design and to the disorder inherent in the construction of randomly-assembled structures. Finally, we compare the networks considered with the real biological networks and find that the latter have in general better performances, due to their higher connectivity, but the former with optimal arrangements can mimic the real networks' behaviour for a specific range of transport parameters. These results show that the use of network-theoretical concepts can be crucial for the characterization and design of efficient artificial energy transport networks.

\end{abstract}

\maketitle

\noindent

\section{Introduction}

Although research on energy transfer in photosynthesis has a very long history~\cite{Photo}, the refinement of experimental techniques in the last decade has dramatically enlarged the range of possible observations at the molecular level~\cite{Structureofmembranes}, thus boosting new interest in the field. Recent progress in this context includes new experimental results aimed at characterizing on one hand the structure of light harvesting systems in other biological organisms \cite{LHCIIstructure} and on the other hand the presence and relevance of quantum effects in the energy transport process \cite{QBioExperiments,QBioTheory}.
These results are of fundamental interest
since they allow to shed light onto the mechanisms at the basis of energy transfer processes, that  owing to the long course of natural selection are likely to be optimally efficient.
Therefore, a more complete understanding of their features has also a potential technological impact in providing useful benchmarks on how to engineer artificial light-harvesting systems. \\
Purple bacteria are among the most important organisms whose photosynthetic apparatus is currently studied. The basic actors in the energy transfer process within these membranes are two kinds of molecular \textit{photosynthetic complexes}, called the \textit{LH1} and the \textit{LH2}. The latter play the role of antennas, capturing the incoming photons and funneling the resulting excitons to the former, that also act as antennas and furthermore contain the \textit{reaction centers} (\textit{RCs}) where charge separation is eventually induced~\cite{Schulten}. \\
The global dynamics in these models is highly dependent on the membrane architecture.
In particular, recent studies have revealed the peculiar capability of these bacteria to adapt the structure of their photosynthetic membranes to the illumination conditions characterizing their growth process~\cite{Sturgis1}.
While theoretical models of the transport process taking place in purple bacteria membranes have already been proposed~\cite{Schulten,Fassioli,Johnson} and they are based on classical, Markovian master equations,  so far only some basic model architectures were considered ~\cite{archi} and a comprehensive study of the relevance of the membrane topology
for efficient energy transfer has yet to be developed.
This may be very important for applications: Current techniques like nanoimprint lithography already allow to form regular patterns of optically active LH2 complexes ~\cite{Escalante} and future developments in this direction might allow to arrange also other complexes, according to more complex designs. \\
In this context, a basic and general question is the following: how is the transport process affected by the topological properties of the underlying structures? Are there design principles allowing to select the most efficient topologies?
In the present work we propose a framework in order to address this questions. We take inspiration from the real biological membranes and simulate the energy transport process ~\cite{Fassioli,Johnson} on more general networks in order to relate the dynamics to different topological properties.\\
Indeed, light-harvesting membranes can be regarded as \textit{networks}~\cite{complex} where the different photosynthetic complexes, such as LH1/LH2, play the role of nodes, and links represent the physical interactions (exciton hopping) between neighbouring complexes. One can thus vary the topological properties of the system by choosing different network architectures and by choosing different arrangements of the complexes on them.
As for the architectures, we will focus on well-known regular structures like Cayley Trees and Regular Hyperbranched Fractals~\cite{Blumen1,vari,Blumen2,DendrimersReviews,Muelken} as well as randomly-generated ones (representing, respectively, presence or lack of control over the networks assemblage).
Once a specific architecture is selected, one has to address the problem of selecting specific arrangements of LH1/LH2 complexes on it. In this  work we propose to use some of the most common network theoretical tools used in the description of dynamical processes on complex networks: \textit{centrality measures}. The latter allow to rank the nodes of a network on the basis of their centrality with respect to it and we will therefore use them as guiding principles for the choice of different arrangements.\\

The results of our analysis show that the topology has a remarkable influence on the dynamics. In particular, for the selected structures, there is a high sensitivity of the efficiency of the energy transport with respect to the choice of the arrangement of photosynthetic complexes.
The use of centrality measures, in particular for regular structures,  allows to highlight specific arrangements that ensure a better efficiency with respect to a random arrangement; we benchmark these results by comparing them with those obtainable by using an optimization algorithm and with those pertaining to the real biological networks.
As for random structures, the efficiency corresponding to a random arrangement is comparable with that achievable with the use of centrality measures; this highlights the robustness of the dynamical process with respect to the lack of control over LH1/LH2 arrangement in randomly-assembled structures.

This work is structured as follows: In Sec.~\ref{Sec : ETpurple} we introduce the basic notions on purple bacteria's light-harvesting membranes and the energy transport model used in the paper. In Sec.~\ref{Sec : networks}  we  introduce the types of networks and the main network-theoretical tools used in this work. In Sec.~\ref{Sec : Results} we  discuss our main results on the relation between network topology and efficiency. Sec.~\ref{Sec : Conclusions} closes the paper with some final remarks.

\section{Photosynthetic energy transfer in purple bacteria} \label{Sec : ETpurple}

\subsection{Structure of the membranes}

\begin{figure}[tb]
\centering
\includegraphics[width=0.4\textwidth]{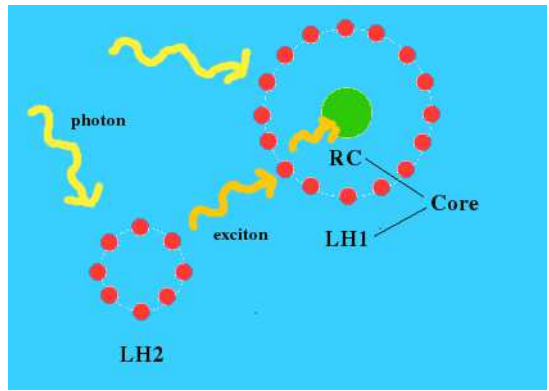}
\caption{Schematic depiction of basic photosynthetic complexes of purple bacteria LH2 and LH1; they both play the role of antennas, capturing the incoming photons. The LH1s also contain the reaction centers (RC) where charge separation eventually occurs.}
\label{Fig.: photomembrane}
\end{figure}

The starting point of our analysis is the energy transport process that occurs in biological organisms such as purple bacteria.
By virtue of techniques like X-ray crystallography and atomic force microscopy, the structure of photosynthetic membranes of purple bacteria has been described with high precision~\cite{Structureofmembranes}; moreover the main dynamical timescales governing the exciton dynamics have been measured or calculated~\cite{transfertimes,Ritz}. The energy transport process is based on two different kinds of pigment-protein \textit{photosynthetic complexes} called \textit{LH1} and \textit{LH2} which grow in the membranes. The \textit{LH2} play the role of antennas, absorbing incoming photons and transferring the resulting electronic exciton to the LH1. The LH1 complexes are also effective in absorbing light, at a different wavelength, and they contain the \textit{reaction centers (RCs)} where the exciton is absorbed by a special pair of clorophylls, inducing ionization of the pair - \textit{charge separation}  - and triggering a chemical reaction (the reduction of quinone to quinol~\cite{Schulten}). \\
LH2 complexes are always more abundant than LH1 ones. However, the \textit{stoichiometry} $s=N_2/N_1$, i.e., the ratio between the number of LH2 ($N_2$) and the number of LH1 ($N_1$), varies depending on environmental conditions. The membranes have a remarkable ability to adapt to the intensity of the illumination during their growth: Bacteria grown under low light intensity (LLI) conditions have a stoichiometry $s_{\lli}\approx 7-9$, while those grown under high light intensity (HLI) conditions have a stoichiometry $s_{\hli}\approx 3.5-5$. As for photon capture the membranes' total cross section is $\sigma=\sigma_1 N_1+\sigma_2 N_2$: since $\sigma_2 = 116 \AA^2 > \sigma_1 = 67.29 \AA^2$, the higher the stoichiometric ratio, the higher the total cross section. The rate of photon capture $R$ is proportional to $\sigma$: $R=I \sigma/h\nu$, where $I$ is the incoming light intensity at the relevant wavelength. As a consequence, under a given illumination instensity $I$, LLI membranes are able to capture more photons than HLI ones, which enables them to survive in scarcer illumination. \par

\subsection{The model of energy transfer}

Excitations are initially created on one complex via photon absorption and can hop to neighboring complexes through a F\"orster resonance mechanism~\cite{Schulten}, until they either dissipate (with typical dissipation time $t_{diss}$) or lead to charge separation in a RC (with typical time $t_{cs}$). A schematic depiction of the process and the basic components is given in Fig.~\ref{Fig.: photomembrane}. A crucial feature of the energy transport process is the fact that when an exciton leads to charge separation within a RC, the latter features a ``busy'' time interval, during which quinol is produced, removed and then a new quinone becomes available. Therefore the RC is \textit{closed} for a time interval $t_{block}$, called \textit{recycling time}, during which it cannot exploit incoming excitons. Furthermore, the RC in the closed state becomes a quencher for excitons, i.e., increases its dissipation rate ($t_{diss}^{RC^c}\ll t_{diss}$).  ). The timescales for energy capture, exciton lifetime and RC reopening are roughly $ R^{-1} \sim 10^{-1}-10^{-2} \ ms, \  t_{diss}= k_{diss}^{-1} \sim 1 \ ns, \ t_{block} \sim 1 ms$ respectively. The recycling time $t_{block}$ can differ significantly depending on the specific kind of bacterion analyzed ($1-30 ms$).
As a consequence of RC closure, at any time only some of the RCs are available for  producing charge separations. The number of available RCs decreases with increasing $t_{block}$ - i.e., as the RCs remain closed for a longer time - and the overall transport process passes from an \textit{active regime} in which all RCs are open to a \textit{saturated regime} in which all RCs are closed. \\
In Table \ref{Tab.: networkparameters} and \ref{Tab.: exciton transfer times} we report the values of the relevant dynamical parameters of the LLI- and HLI- adapted membranes~\cite{transfertimes,Fassioli,Johnson}.

The problem of modeling energy transfer processes in biological and artificial systems has been thoroughly studied in the past decades, and models of the exciton transport dynamics in light-harvesting membranes have been developed in order to take into account both incoherent and coherent phenomena~\cite{OlayaScholesETreview}. Here we focus on a simple \textit{classical, Markovian model} developed in~\cite{Fassioli,Johnson} to model energy transfer in purple bacteria membranes. Since exciton transfer between complexes in such membranse arises through the Coulomb interaction on the $ps$ time-scale, while vibrational dephasing destroys coherences within a few hundred $fs$, coherent effects in the energy transfer between complexes are expected to be very weak and are therefore neglected in these models~\footnote{While it might be of interest to investigate the possible role of weak quantum effects, we neglect this issue since the focus of our paper is on the relation between transport and topology and a comparison between quantum and classical transport models is beyond its scope.
While noiseless quantum models on dendrimers exhibit the tendency to localization of the exciton on some specific (outermost) nodes of the network (see e.g.~\cite{Muelken}), in more realistic quantum models one should take into account the role of dephasing and/or dissipation, allowing for the appearance of delocalization of the exciton and environment assisted energy transport.}\\
Since $t_{diss}\gg R^{-1}$, simultaneous occurrence of two excitons in the membrane is very unlikely, therefore the system can be modeled as if \textit{a single exciton} were present at each time $t$.
The dynamics is described by a Markovian Master equation (see appendix B for details) and thus can be numerically simulated by standard random walk methods~\cite{Fassioli,Johnson}. In the following, we will always base our analyses on random walk simulations. \\
Excitons are created at random times $\{t_i\}$ determined in advance by using a Poissonian distribution. At each time $t_i$ a single exciton is randomly created in a LH1 (LH2) site with probability $p_1=\frac{N_1 \sigma_1}{N_1 \sigma_1 +N_2 \sigma_2}$ $(p_2=1-p_1)$. The exciton then follows a random walk, and the probability of jumping from a site $j$ in a given time step $\delta t$ is given by $p^j_{jump}(\delta t)=K_j \delta t$ with
\be
K_j=\sum_{i\neq j}W_{ij}+k_{diss}(1-\delta_{j,RC^c})+k^*_{diss}\delta_{j,RC^c}+k_{cs}\delta_{j,RC^o}
\ee
where $k_{diss}=t_{diss}^{-1}$, $k^*_{diss}=t_{RC^c_{diss}}^{-1}$, $k_{cs}=t_{cs}^{-1}$ and the exciton transfer rates $W_{ij}$ are zero if the nodes are not linked, else they are taken as the inverse of the typical exciton transfer times $t_{X \to Y}$. The values of all timescales are the same as in Tables \ref{Tab.: networkparameters} and \ref{Tab.: exciton transfer times}. The excitons remain on the same site or jump to a neighboring one according to different probabilities, until they dissipate or lead to charge separation in a RC. When charge separation occurs in a RC, the latter  remains closed for time $t_{block}$ in the simulation and its dissipation rate is increased.
Our simulations cover a long time of $\sim 1 s$, so that the histories of $N_{abs}=10^5$ excitons are simulated \cite{notealgorithm}. \\
The optimality and robustness of energy transfer are characterized by performance measures: The most relevant is the  efficiency of the transport process $\eta$, i.e, the probability that an exciton lead to a charge separation - while $1-\eta$ is the probability that it dissipate. In the random walk,
we evaluate the ratio $\eta = N_{cs}/N_{abs}$, where $N_{cs}$ is the number of photons that are used for charge separation in RCs and $N_{abs}$ the total number of photons absorbed by the network.

\begin{table}[tb]
\caption{Relevant scale parameters for HLI and LLI networks: stoichiometric ratio, absorption cross-section, and dynamic timescales}
\centering
\begin{tabular}{|c|c|c|c|c|c|c|c|}

  \hline
      & I & $N_{LH2}/N_{LH1}$& $\sigma$ & $R^{-1}$ & $t_{diss}$ &  $t_{diss}^{RC^c}$& $t_{cs}$\\  \hline
  HLI & $100 \, w/m^2$ & $4.64$& $\approx 12 {\AA}^2$ & $0.02\, ms$ &$1\, ns$ & $30\, ps$ & $3\, ps$\\  \hline
  LLI & $10 \, w/m^2$ & $7$ &$\approx 14 {\AA}^2$& $0.16\, ms$  & $1\, ns$ & $30\, ps$& $3\, ps$ \\  \hline
\end{tabular}
\label{Tab.: networkparameters}
\end{table}
\begin{table}[bt]
\caption{Typical transfer times between different photosynthetic complexes}
\centering
\begin{tabular}{|c|c|c|c|c|c|}
  \hline
  $t_{\scriptscriptstyle LH1 \rightarrow LH1}$  &
  $t_{\scriptscriptstyle LH2 \rightarrow LH2}$ &
  $t_{\scriptscriptstyle LH2\rightarrow LH1}$  &
  $t_{\scriptscriptstyle LH1 \rightarrow LH2}$ &
  $t_{\scriptscriptstyle LH1 \rightarrow RC}$  &
  $t_{\scriptscriptstyle RC \rightarrow LH1}$  \\ \hline
  $20\, ps$ & $10\, ps$ & $3.3\, ps$ & $15.5\, ps$ & $25\, ps$ & $8\, ps$ \\
  \hline
\end{tabular}\label{Tab.: exciton transfer times}
\end{table}

\section{Biologically-inspired networks and network-theoretical tools} \label{Sec : networks}


In order to study the role of topology for the biologically inspired transport process described in the previous section, we will consider different kinds of regular and randomly generated networks. The structures that we are going to use are based on a pair $\{\Gamma, \mathcal{A} \}$ composed by a graph $\Gamma=(E,V)$, with set of  vertices $V$ and  edges $E$, together with an arrangement $\mathcal{A} = (V_1, V_2 )$ i.e., a subdivision of the vertices in two \textit{disjoint} subsets $V_1, V_2 \subset V $ that identify the nodes that are occupied by LH1 and LH2 respectively.
The complete structure is obtained by adding to $\Gamma$ one node for each element of $V_1$, representing a RC directly linked only to the respective LH1 complex.
We have  $|V_1|=N_1$ and $|V_2|=N_2$, where the number of the different complexes is determined by the stoichiometry$\ s$. We will investigate two different settings, representing specific high light intensity (HLI) and low light intensity (LLI) conditions. The stoichiometric ratios are those found in real networks: $N_2/N_1=4.64$ for for HLI,  $N_2/N_1=7.04$ for LLI. \\\\
In general, the transport process properties will be determined both by the topological features of the graphs $\Gamma$ and by choice of the arrangement $\mathcal{A}$, i.e, how the LH1/LH2 complexes are placed on $\Gamma$.
The different arrangements $\mathcal{A}$ will be chosen following criteria based on tools developed in complex networks theory (centrality measures).
The selected structures will be compared with the real networks describing purple bacteria membranes, from the point of view of energy transfer properties.

\subsection{Networks} \label{Subsec : selected networks}

\begin{figure}[bt]
\includegraphics[width=0.35\textwidth]{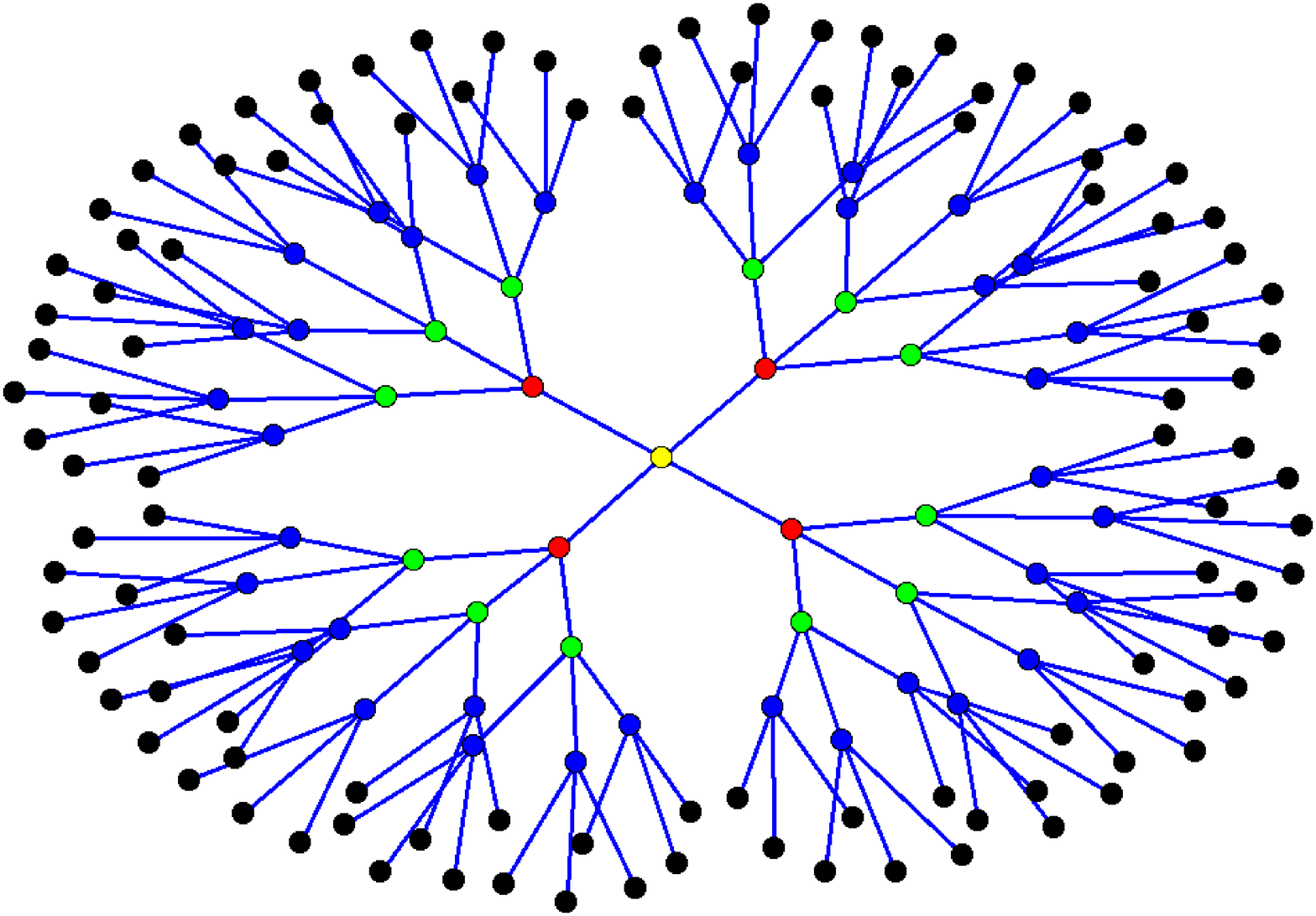}
\includegraphics[width=0.45\textwidth]{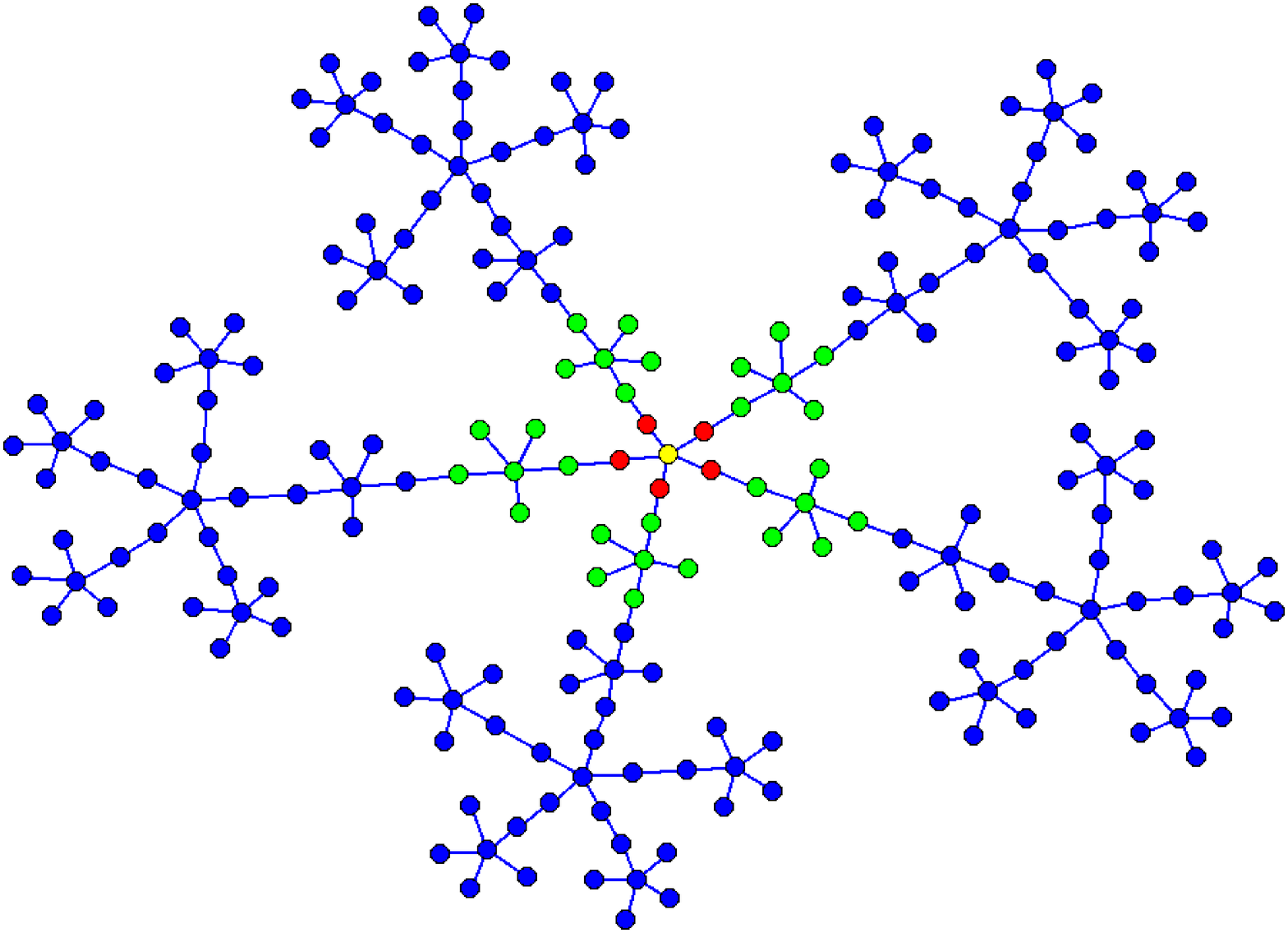}
\caption{Regular graphs, generated according to the procedure in subsection \ref{Subsec : selected networks}: (Top) Cayley tree with $d=4$, $n=4$ (Bottom) RHF with $f=5$, $n=3$. In both graphs, nodes of different colors belong to different generations: 0 (yellow), 1 (red),  2 (green),  3 (blue), 4 (black).} \label{Fig.: regulargraphs}
\end{figure}

\begin{figure}[hbt]
\includegraphics[width=0.4\textwidth]{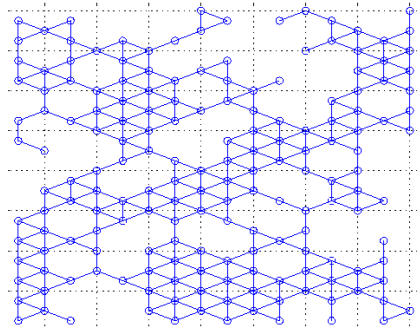}
\caption{Random network generated according to the procedure in subsection \ref{Subsec : selected networks}}  \label{Fig.: randomgraphs}
\end{figure}

The regular networks we will focus on are well-known regular structures which have been considered ~\cite{Blumen1,vari,Blumen2} in models of energy transport: $i)$ { \it Cayley trees}  (CT) $ii)$ {\it regular hyperbranched fractals} (RHF). \\
The CT are constructed as follows. The $n$-th generation $d$-Cayley
tree  is a tree of $n$ levels in which all vertices on the interior have
degree $d$, while vertices on the outermost layer have degree 1. In our simulations we use $d=4,n=4$ (see fig.~\ref{Fig.: regulargraphs}) and a total number of nodes is $161$.\\
The RHF instead are constructed as follows. A first generation (RHF) (as shown in fig.~\ref{Fig.: regulargraphs}) of functionality $f$ is a star graph consisting of a central vertex connected through $f$ edges to $f$ surface vertices. To construct a second generation RHF, $f$ copies of the first generation RHF are
connected to the core first generation RHF through a single leaf-leaf edge. This procedure
is repeated $n$ times for an $n$-th generation RHF. In our simulations we use $f=5, n=3$ and the total number of nodes is $216$.
Both CT and RHF depicted in fig. \ref{Fig.: regulargraphs}. \par
In addition to regular structures, we also investigate randomly-generated ones. This allows to compare
the effiency of transport achieved in two very different cases: the case where one has global control over the network topology $\Gamma$ and the case where one has no control on it - this might happen, for instance, if the network is randomly assembled.
A  natural way to generate planar structures that are possible realizations of light-harvesting networks is to consider for example {\it Randomly-decimated hexagonal Networks} (RN). These are obtained according to the following prescription: We consider a hexagonal lattice of $N_{tot}=256$ sites and remove $m$ randomly-chosen links such that the connectedness of the network is preserved. In our simulations $m=193, 321$ so that the average degree becomes $\sim 4$ and $\sim 3$ respectively. For both values of $m$ we generate an ensemble of $M=100$ RNs (a particular element of the ensemble is depicted in fig. \ref{Fig.: randomgraphs}).
\par

\begin{figure}[bt]
\centering
\includegraphics[width=0.3\textwidth]{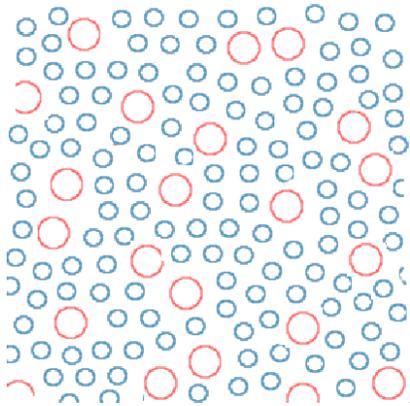}
\caption{Simplified depiction of a typical HLI photosynthetic membrane of purple bacteria}
\label{Fig.: photomembrane2}
\end{figure}

Finally, we consider the \textit{real, biological networks} studied by Fassioli et al.~\cite{Fassioli}, which are based on studies of the \textit{Rsp. photometricum} membranes~\cite{Sturgis1}. The global membrane structure is known in great detail, since atomic force microscopy has provided high-resolution images of the membranes of the purple bacterium \textit{Rsp. photometricum}~\cite{Sturgis1}.  The membranes have a planar structure where complexes are densely packed, and the cores are surrounded by LH2 complexes (5-7 on average)~\cite{Sturgis1}. A simplified depiction of a typical HLI membrane in terms of the constituent complexes is given in fig. \ref{Fig.: photomembrane2}; in the network model, links are present between neighbouring complexes.
Notice that we consider a single realization of real network for HLI and LLI i.e., in both cases $\Gamma$ and $\mathcal{A}$ as fixed and correspond to the networks in ~\cite{Fassioli}.
Furthermore, LLI and HLI membranes, considered as networks, have a different topology: since LH1 have on average more links to neighbouring complexes than LH2, HLI structures have a higher average connectivity. The total number of nodes are $168$ (HLI) and $193$ (LLI). \par
The parameters defining all the above networks have been chosen in order to have structures with approximately the same size i.e., the same number of nodes $|V|$.

\subsection{Choosing different arrangements: Centrality measures}

Once a specific network $\Gamma$ is chosen, another step is needed in order to completely specify the topology of the transport network. Indeed, different arrangements $\mathcal{A}$ of the LH1/LH2 complexes are possible and one has therefore to identify guiding principles allowing for the selection of different inequivalent arrangements. In this sense, a possibility is given by the study of the relevance of the nodes within a network. Great efforts in the development of the theory of complex networks have been devoted to this task and in this context several \textit{centrality measures} have been proposed~\cite{Borgatti}. The usefulness of a given measure depends strongly on the specific problem at hand, since the latter determines which nodes are more relevant from case to case.\\
Two common families of centrality measures are the set of degree-like and the set of betweenness-like centrality measures.
Degree-like measures are essentially based on counting the number of paths emanating from the node. Different measures are obtained considering different path lengths and different kinds of paths (edge- of node- disjoint, geodesic, etc.). The simplest of such measures is the \textit{degree}(DEG)~\cite{Freeman1}: The degree $k$ of a node is simply the number of nodes to which it is connected (i.e., the number of paths of length 1 emanating from the node).
Betweenness-like measures instead are based on counting the number of paths that pass through a given node $k$. Again, different measures are obtained considering different path lengths and different kinds of paths. The most common of such measures is the \textit{shortest path betweenness centrality}(SPB)~\cite{Freeman2}; here the basic objects are the shortest paths connecting two nodes $i,j$ (a path is given by a sequence of edges connecting the two nodes) ; in general there are many different paths of minimal length (sequences with minimal number of edges) between two nodes. The SPB of a node $k$ is defined as $B_k = \sum_{ij} \frac{g_{ijk}}{g_{ij}}$ where $g_{ij}$ is the total number of shortest paths from node $i$ to node $j$, while $g_{ijk}$ is the subset of such paths passing through node $k$.
While the degree is a local measure of node centrality, depending only on the structure of the network in the immediate neighbourhood of a given node, the shortest-path betweenness is a global measure of node centrality, since it takes into account the structure of the whole network. \par
By means of the above measures one can identify different inequivalent arrangements $\mathcal{A}$ in the following way.
For each $\Gamma$ we evaluate the DEG, SPB of all nodes  (a full characterization of $\Gamma_{CT}, \Gamma_{RHF}$ and $\Gamma_{RN}$ in terms of these measures is presented in the Appendix A). An arrangement $\mathcal{A}$ is identified by choosing the subset $V_1$ as the set of $N_1$ nodes satisfying one of the following criteria\footnote{If a given criterion does not allow to univocally identify the core position on the network we randomize the corresponding possible choices.}
\begin{itemize}
 \item nodes with maximal SPB, or DEG
 \item nodes with minimal SPB, or DEG
\item nodes chosen at random
\end{itemize}

While other choices are possible, these are the basic and natural ones that allow to test how the transport efficiency can be affected by different arrangements of the RCs on a given network. Like in the case of random topology, the random arrangement represents the case when no control over the LH1/LH2 arrangement is possible.

\section{Energy transport in biologically inspired networks} \label{Sec : Results}

In the following we give a detailed description of the dynamics and then we discuss the relation between the dynamics and the topological features of the selected structures.
In order to describe the dynamics we introduce the following functions of the recycling time $t_{block}$:
\begin{itemize}
\item{} the \textit{efficiency} $\eta = N_{cs}/N_{abs}$;
\item{}the \textit{average fraction of closed RCs} $\average{n_{RC}^c}=\average{N_{RC}^c/N_{RC}}$;
\item{}the \textit{average lifetime} $\average{\tau}_{cs}$ of successful excitons, i.e., those that reach a RC and induce a charge separation;
\item{}the \textit{average maximal distance} $\average{d}_{cs}$ of successful excitons, i.e., the distance, in steps, between the initial site and the most distant site reached during the random walk);
\item{}the \textit{average exploration parameter} $\average{x}_{cs}$ of successful excitons, i.e., the fraction of network sites visited.
\end{itemize}
In particular $\average{d}_{cs}$ and $\average{x}_{cs}$ characterize the diffusion of excitons over the network.
\\
For each choice of $\Gamma$ and $\mathcal{A}$, and for any fixed value of $t_{block}$
we realize the random walk dynamics corresponding to $N_{abs}=10^5$ excitons. The above averages are taken over the $N_{abs}$ histories
($\eta, \average{n_{RC}^c}$) or over the histories of the $N_{cs}$ successful excitons ($ \average{\tau}_{cs}, \average{d}_{cs}, \average{x}_{cs}$).
In case of random networks or arrangements the functionals are the results of a further average over the elements of a given ensemble of networks and/or arrangements.
Finally we let  $t_{block}$ varying over a large range of values ($10^{-3} ms - 10^{3} ms $).

\subsection{ Cayley trees}

\begin{figure}[bt]
\begin{tabular}{c}
\subfloat[]{\includegraphics[width=0.8\columnwidth]{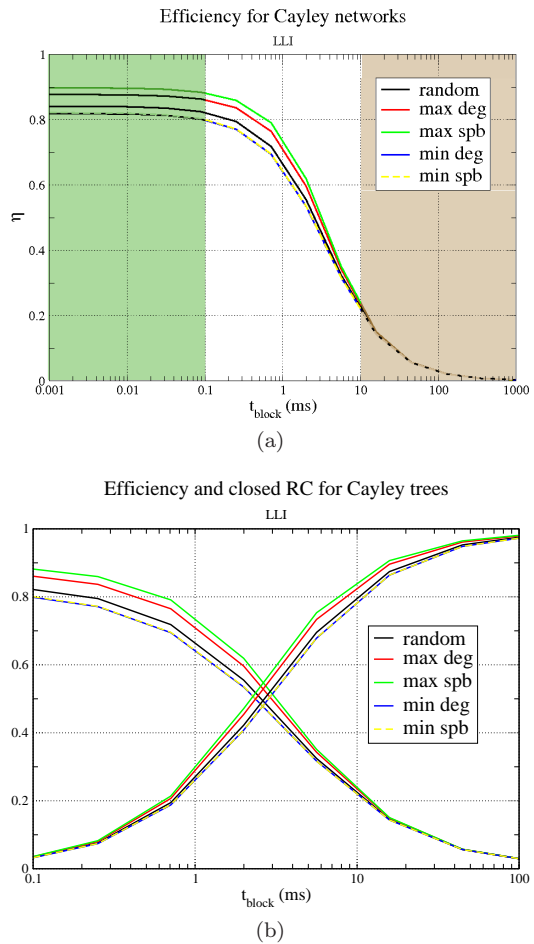}}\\
\subfloat[]{\includegraphics[width=0.8\columnwidth]{x_Cayley_low_final.eps}}
\end{tabular}
\caption{ (Top) Efficiency $\eta$ of exciton transport for CT as a function of $t_{block}$ in LLI conditions, under different LH1 arrangements. One can identify the active (shaded green) and saturated (shaded red) regimes. The transition occurs at $t_{block} \sim R^{-1} \sim 0.1 $ ms. (Bottom) $\eta$ and RC occupation  $n_{RC}^c$ for CT as a function of $t_{block}$ in LLI conditions. $\eta$ sinks and $n_{RC}^c$ grows when $t_{block}$ is increased. For any fixed $t_{block}$, efficient arrangements correspond to higher values of both $\eta$ and $n_{RC}^c$. }
\label{Fig.: efficiency_Cayley_Intensity}
\end{figure}

We now show how the introduced functionals allow to characterize the dynamics and its relation with the topology of LH1/RC arrangement. 
We shall first focus on CTs, and then analyse results for the remaining network topologies (RHF, RN). \par
{\it Different transport regimes.} As a first general comment, we note that the dynamics is characterized by three main time scales, given by the absorption rate $R^{-1}$, the typical dissipation time $t_{diss}$  and the recycling time $t_{block}$.  In particular there are two distinct regimes: The \textit{active} regime, characterized by  $ t_{block}\, < \, R^{-1} \, \ll\ t_{diss}$, i.e., by a recycling time smaller than the absorption rate; and the \textit{saturated} regime, characterized by  $ R^{-1}\, < \,t_{block}\, \ll\,t_{diss} $, i.e., by a recycling time greater than the absorption rate. The transition between the two regimes occurs at $R^{-1}\, \approx \,t_{block}$. \\
In fig.~\ref{Fig.: efficiency_Cayley_Intensity} (upper panel)
we show $\eta$ for CTs for all different LH1 arrangements in LLI conditions. The behaviour for HLI networks (not shown) is analogous; since $R^{-1}_{HLI}\approx 10^{-1} R^{-1}_{LLI}$ the transition between the two regimes occurs in HLI networks earlier than in LLI ones.
\par
Each absorbed photon moves in a network which is characterized not only by the stoichiometric ratio, but also by the number of closed RCs. In fig.~\ref{Fig.: efficiency_Cayley_Intensity} (right panel) both $\eta$ and $n_{RC}^c$ are plotted together for the different LH1/RC arrangements. On one side, for any given arrangement,  $n_{RC}^c$ grows with increasing $t_{block}$, while $\eta$ sinks.
On the other side, if we compare different arrangements at a fixed value of $t_{block}$, we notice that more efficient arrangements have a larger fraction of closed RC: excitons are more likely to be absorbed, so that on average more RCs are closed and consequently the ordering of the efficiency curves is the same as the RC-closure curves.\par
\begin{figure}[htb]
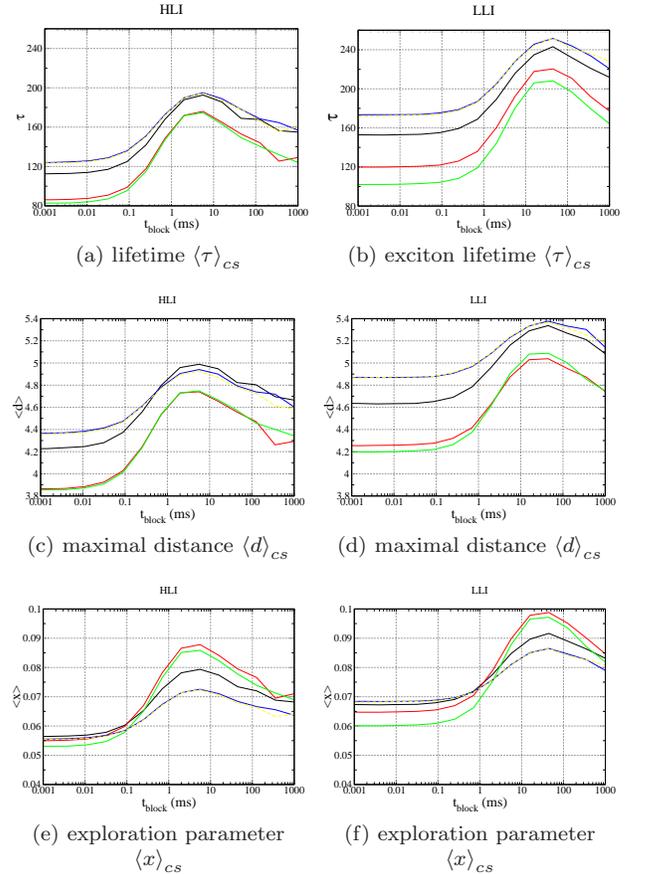

 \begin{tabular}{cc}
  \subfloat[lifetime $\average{\tau}_{cs}$]{\label{Fig.: tau_Cayley_high} \includegraphics[width=0.45\columnwidth]{tau_Cayley_high_final.eps}} &
  \subfloat[exciton lifetime $\average{\tau}_{cs}$]{\label{Fig.: tau_Cayley_low}
  \includegraphics[width=0.45\columnwidth]{tau_Cayley_low_final.eps}} \\
  \subfloat[maximal distance $\average{d}_{cs}$]{\label{Fig.: distance_Cayley_high}\includegraphics[width=0.45\columnwidth]{distance_Cayley_high_final.eps}} &
  \subfloat[maximal distance $\average{d}_{cs}$]{\label{Fig.: distance_Cayley_low}\includegraphics[width=0.45\columnwidth]{distance_Cayley_low_final.eps}} \\
  \subfloat[exploration parameter $\average{x}_{cs}$]{\label{Fig.: exploration_Cayley_high}\includegraphics[width=0.45\columnwidth]{exploration_Cayley_high_final.eps}} &
  \subfloat[exploration parameter $\average{x}_{cs}$]{\label{Fig.: exploration_Cayley_low}\includegraphics[width=0.45\columnwidth]{exploration_Cayley_low_final.eps}}
\end{tabular}
\caption{Lifetimes $\average{\tau}_{cs}$, average maximal distances $\average{d}_{cs}$, exploration parameter $\average{x}_{cs}$ of successful excitons as a function of the recycling time $t_{block}$ for Cayley trees in HLI (left) and LLI (right) conditions, under different LH1 arrangements: (black) random (red) max DEG (green) max SPB (blue) min DEG (yellow) min SPB. All quantities grow in the transition region, due to the necessity of longer paths in
order to find open RCs, and have larger values for LLI networks that have less RCs.
\label{Fig.: exploration_Cayley}
}
\end{figure}

As for the other functionals, $\average{\tau}_{cs}, \average{d}_{cs}, \average{x}_{cs}$ we note that they all display the same behaviour with $t_{block}$: they remain essentially constant in the active regime and they grow throughout the transition region. Indeed when many RCs are closed, excitons are expected to travel further away from their initial site and explore a larger fraction of nodes in order to find open RCs. A maximum is reached when $n_{RC}^c \sim 90\%$. Thereafter, unless one of the very few open RCs is found in the vicinity of the initial exciton site, the excitons are dissipated before reaching any RC and thus all quantities undergo a slight decrease after this threshold. \\
The concentration of RCs is also at the basis of the difference between LLI and HLI networks. Indeed, the values of  $\average{\tau}_{cs}, \average{d}_{cs}$ and $\average{x}_{cs}$ are higher in the former case; due to lower stoichiometric ratio, the excitons have to explore a bigger fraction of the network; for instance, in the active photosynthesis regime one has that for LLI $\average{\tau}_{cs} \in [ 90 ps,180 ps ]$ while for $\average{\tau_{cs}} \in [ 80 ps,120 ps ]$ for HLI networks.\\
The plots of $\average{x}_{cs}$ in Fig.~\ref{Fig.: exploration_Cayley} (panels $c$ and $f$) show that the absorbed excitons do not explore the whole network, but only a relatively small portion of it: only $ 5-10\%$ of all sites are visited by successful excitons.
Furthermore, the ordering of the $\average{d}_{cs}$ and $\average{\tau_{cs}}$ curves is reversed with respect to the $\eta$ ones.
We thus reach a first important conclusion: due to dissipation effects, excitons cannot explore the whole network, and in efficient configurations excitons find open RCs in the vicinity of the initial site and reach it in the shortest time. 
\textit{Optimal arrangement criterion and topological properties of the network.}
In the case of the Cayley tree the LH1 arrangement criterion giving the highest efficiency is the maximal SPB, and the corresponding subset $V_1$ is composed by the nodes clustered around the root of the network. This fact can be explained by analyzing the topological properties of the network $\Gamma_{CT}$ and the dynamical behaviour.
Indeed, in the active photosynthesis region the distance $\average{d}_{cs}\approx 4$ is comparable with a key topological property of the network, namely, the average distance between two nodes of the network which is of $6.4$ steps.
This means that, although the exploration parameter is relatively small, the exciton is able to move across the network e.g., from the periphery to the center of the tree and vice-versa. Therefore, it can \textit{take advantage of the global character of the centrality} of the nodes occupied by the LH1/RCs.
This is confirmed by the positive value of the linear correlation coefficient between the SPB of each node and the average time spent on it, which we find to be approximately $0.7$ in all regimes: excitons are likely to reach the central region and to spend more time on the $V_1$ nodes.\\
A general important feature highlighted by our analysis is that both for LLI and HLI structures the maximal SPB criterion allows for a \textit{significant improvement with respect to the random arrangement}. This indicates that the chosen network theoretical tools represent a \textit{useful guiding principle} for the identification of the most efficient configurations.

\begin{figure}[tb]
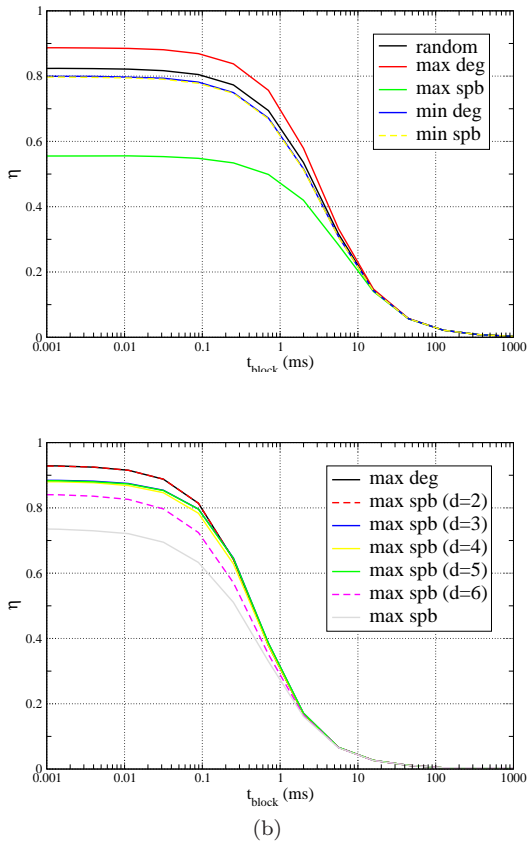

\centering
\begin{tabular}{c}
\subfloat[]{\includegraphics[width=0.8\columnwidth]{efficiency_RHF_low_final.eps}} \\
\subfloat[]{\includegraphics[width=0.8\columnwidth]{efficiency_RHF_high_comp_final.eps}}
\end{tabular}
\caption{(Top) efficiency curves for RHF at LLI under different LH1 arrangements. The arrangement based on global SPB is particularly inefficient. (Bottom) efficiency curves for RHF at HLI. Arrangements are based on max $SPB(d)$, $d=2-6$);  efficiency grows as an increasingly local structure of the network is considered.}
\label{Fig.: efficiency_RHF_Intensity}
\end{figure}

\subsection{Regular Hyperbranched Fractals.} In the case of RHF networks,
the global behavior of all functionals with $t_{block}$ is analogous to the one described for the CTs. However, some fundamental differences emerge.
On one hand, the range of the various functionals is much higher for the RHF and this is a clear indication that these networks are very sensitive to the choice of the LH1 arrangement. For example, in the active photosynthesis regime $\average{\tau}_{cs} \in [ 100 ps, 375 ps ]$ for LLI while for HLI $\average{\tau}_{cs} \in [ 70 ps,240 ps ]$.\\
On the other hand, the criterion that allows for a maximal efficiency is in this case the maximal DEG,
i.e., a \textit{local} measure of centrality of the nodes, wheareas the arrangement determined by the maximal SPB criterion (Fig. \ref{Fig.: efficiency_RHF_Intensity}, left panel) has a considerably low efficiency ($60\%$).
This result can again be explained in terms of the topological features of the $\Gamma_{RHF}$.
Indeed, the values of $\average{d}_{cs}$ are in general small if compared to the average distance between the nodes of the $\Gamma_{RHF}$ network which is $16.68$ steps (for example $\average{d}_{cs} \in [4, 6]$ for LLI case in the active photosynthesis region).
Therefore, the dynamics \textit{can not exploit the global centrality} of the nodes with max SPB i.e, those clustered around the root of the network;
This fact is again confirmed by the linear correlation coefficient between the betweenness of each node and the average time spent on it which is now negative and equal to  $-0.5$ in the case of max SPB arrangement.\\
The previous result suggests a general consideration: for networks with high values of the average distance between nodes and in presence of a dissipative dynamics that does not allow for high values of $\average{d}_{cs}$, optimal LH1 arrangements can be found by considering local centrality  measures, able to characterize the centrality of the various nodes with respect to the \textit{local surrounding network}.
In order to further test this fact we introduce a class of centrality measures which generalize the concept of local betweenness centrality already introduced in the context of search problems on complex networks~\cite{Thadakamalla}. For each node $k$ of a given network $\Gamma$, one can fix a distance $d$ and consider the subnetwork composed by those nodes whose distance from $k$ is $\leq d$. One then evaluates the shortest path betweenness of the given node in the selected subnetwork. We shall denote such ``d-local'' betweenness centrality measure as $SPB(d)$~\cite{NoteSPB(d)1}. The latter coincides with the SPB for sufficiently high values of $d$). For each value of $d$ the arrangement we consider corresponds to placing the LH1 complexes on the nodes with higher values of $SPB(d)$.
Let us consider RHFs and compare the results obtained with the LH1 arrangement based on global SPB and that based on $SPB(d)$ for various values of $d$. In Fig.~\ref{Fig.: efficiency_RHF_Intensity}(right panel) we clearly see, that while the global SPB induces a very inefficient arrangement, the local $SPB(d)$ measures progressively induce more efficient arrangements as $d$ decreases and a increasingly local structure of the network is considered \cite{NoteSPB(d)2}. Even though this precise hierarchy is specific to the RHF, it is also a good qualitative example of how local, rather than global centrality measures may be more appropriate for dissipative dynamics that take place on networks with large average distance between sites.\par
We conclude our discussion by observing that also in the case of RHF structures, the use of criteria based on centrality measures allow to identify optimal arrangements whose efficiencies are significantly higher than those obtained by randomly placing the various complexes.

\subsection{Random networks}

\begin{figure}[tb]
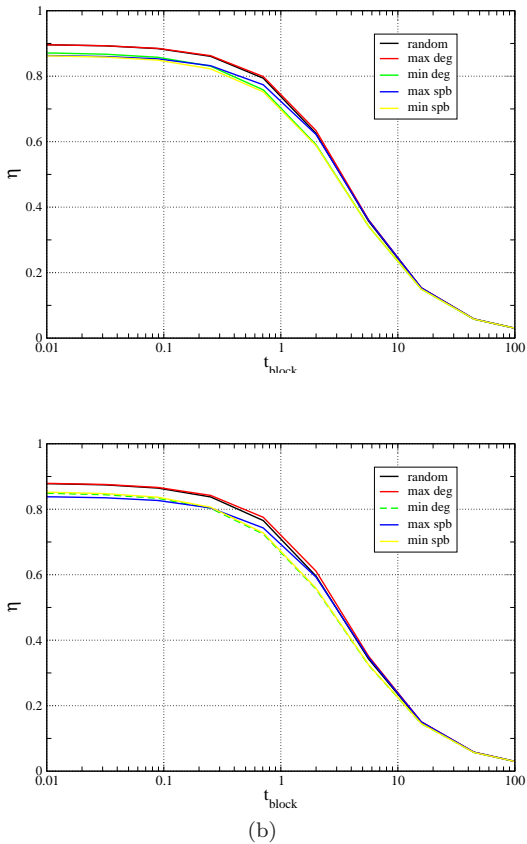

\begin{tabular}{c}
  \subfloat[]{\label{Fig.: efficiency_random_193}
\includegraphics[width=0.8\columnwidth]{efficiency_random_link_193_low.eps}} \\
  \subfloat[]{\label{Fig.: efficiency_random_321}
\includegraphics[width=0.8\columnwidth]{efficiency_random_link_321_low.eps}}
 \end{tabular}
\caption{efficiency for random networks at LLI under different LH1 arrangements: (Left) $m=193$ (Right) $m=321$. A lower connectivity leads to a slightly lower efficiency.
The best topological criterion (max DEG) has approximately the same efficiency of the random arrangement.}
\label{Fig.: efficiency_random_link}
\end{figure}
We now pass to examine the random networks generated according to the procedure described in Sec. \ref{Sec : networks}. Notice that in the following discussion and figures, we always cosider averaged quantities over an ensemble of $M=100$ randomly generated networks.
In Fig.~\ref{Fig.: efficiency_random_link} we show the efficiency curves for the LLI networks with $m=193,321$ links eliminated (average DEG$=3,4$ respectively).

The efficiency $\eta$, as well as the other functionals, globally follow the same behavior of the previously examined cases.
The main difference between the two cases $m=193$ and $m=321$ is the global connectivity which is higher in the former case. This allows for higher values of $\average{d}_{cs}$ and $\average{x}_{cs}$ and in general for slightly higher values of efficiency.\\
In both cases, the criterion that gives the best efficiency is a local one, i.e., the maximal DEG.
As for the relation between topology and arrangement criteria, we first focus on the LLI case with $m=321$, which are the networks which show the maximal variation of the various functionals with respect to the change of the arrangement. Here the average maximal distance traveled in the saturation regime is $\average{d}_{cs}\in [5.8,6.5]$, which has to be compared with the average distance between pairs of sites which is $\approx 11$. This 
is consistent with the fact  that the criterion giving the best efficiency is a local one, i.e., the maximal DEG.
However, for the case $m=193$, the values of $\average{d}_{cs}$ are in general higher than the $m=321$ case while the average distance between pairs of sites is smaller ($\approx 9$), which would suggest a global criterion like max SPB. Therefore we have an indication that in the case of randomly generated networks these two parameters are not sufficient to identify an optimal arrangement criterion. \\
Moreover, at variance with the case of regular networks, for all values of $t_{block}$ the efficiency corresponding to the best topological criterion (max DEG) is only slighty better than that obtained with a random arrangement of the LH1s. This result is very relevant since it suggests that the dynamical process (originally modeled for purple bacteria membranes) is a sense optimized to take into account the disorder inherent in the construction of randomly-assembled structures. In particular, this should hold for the real biological structures, but also for artificial structures where no control on the network structure $\Gamma$ is possibile.
Global optimization of the topology is indeed a rather difficult task (biologically and artificially). Therefore, on one hand a dynamical process like energy transfer has to be strong with respect to limited amount of control in the structure ($\Gamma$, $\mathcal{A}$) design, in order to be successful. On the other hand, if at least a local control in the topology of the network is possible (this should be in principle an easier task to accomplish), then the process could be devised in order to take advantage of the max DEG arrangement.

\subsection{Sensitivity of the efficiency to arrangement criteria in different network topologies}

As we have noticed, different LH1/RC arrangements can cause significant differences in efficiencies, lifetimes and all other functionals. In the following we shall focus on $\eta$, which is by far the most important one - being the main goal of biological and/or artificial optimization.
In particular, the basic feature we want to study is the \textit{sensitivity of the efficiency} of a given network with respect to the choice of the LH1s/RCs arrangement. In order to quantify this sensitivity we introduce the functional $\delta \eta=|\eta_M-\eta_m|/\eta_M$ where, for any fixed values of $t_{block}$, $\eta_M$ ($\eta_m$) is the maximal (minimal) efficiency corresponding to the optimal $\mathcal{A}_M$ (worst, $\mathcal{A}_m$) arrangement.

\begin{figure}[htb]
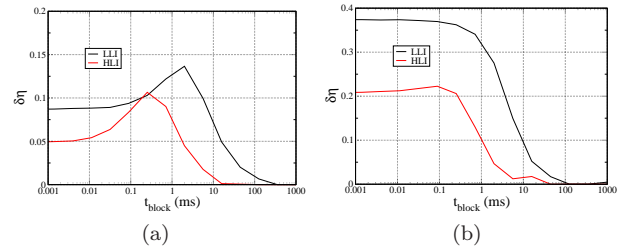

  \centering
 \begin{tabular}{ccc}
  \subfloat[]{\label{Fig.: deltae_Cayley}
  \includegraphics[width=0.45\columnwidth]{deltae_Cayley_g.eps}} &
  \subfloat[]{\label{Fig.: deltae_RHF}
  \includegraphics[width=0.45\columnwidth]{deltae_RHF_g.eps}}
 \end{tabular}
  \caption{$\delta \eta$ vs $t_{block}$ for the Cayley trees (left) and RHF (right). In LLI networks the LH1 fraction is lower, so networks are more sensitive to LH1 arrangement}
  \label{Fig.: deltaeta RN_Cayley_RHF}
\end{figure}

In fig.~\ref{Fig.: deltaeta RN_Cayley_RHF} the sensitivities of CTs and RHFs are shown as a function of $t_{block}$. A first important feature is that the LLI networks are in general more sensitive than the HLI ones.
Thus, for higher values of the stoichiometric ratio, the definition of criteria for choosing RCs arrangements becomes crucial in order to have efficient energy transport. \\
As for CTs the sensitivity has some initial value ($5-9\% $) in the active regime, it grows to a maximum ($\approx 12 - 14\% $) approximately when $t_{block}=\bar{t}\approx 2 R^{-1}$ and then falls off to zero when $t_{block}\gg R^{-1}$. It is therefore in the transition regime that in general the energy transport is most affected by the choice of RCs arrangement.

\begin{figure}[htb]
\includegraphics[width=\columnwidth]{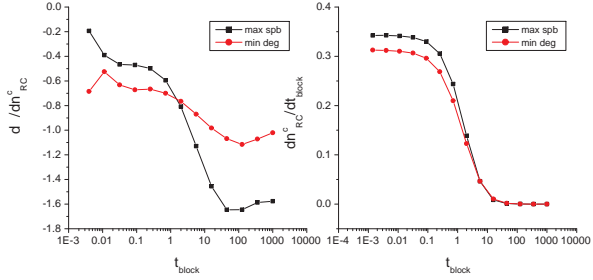}
\caption{Relation between $\Delta \eta$ and $n_{RC}^c$ for Cayley trees at LLI:  (left) $d\eta/d n_{RC}^c$ (right)  $\dot{n}_{RC}^c$}
\label{Fig.: analysis_Cayley}
\end{figure}

This feature is rooted in the ability of most efficient configurations to ensure high value of efficiency and at the same time a higher \textit{resilience} to RCs closure. In order to clarify this point
we focus on the absolute sensitivity $\Delta \eta=\eta_M-\eta_m$ (here $M=$ maximum SPB, $m= $ minimum degree), which mirrors the behavior of $\delta\eta$, and on its derivative with respect to the recycling time $t_{block}$; in the following we define $d \eta/ d t_{block}\doteq \dot{\eta}$. In the initial part of the transition region ($t_{block} < \bar{t} $) we have that $\dot{\eta}_m<\dot{\eta}_M<0$ , i.e., the efficiency for the best arrangement $\mathcal{A}_M$ \textit{decreases less} than that of the least efficient one $\mathcal{A}_m$  and therefore  $\Delta \dot{ \eta}>0$. The derivative of the efficiency with respect to $t_{block}$ can be decomposed in the product of two contributions $\dot{\eta}= \dot{n}_{RC}^c \,\,\, d \eta/ d n_{RC}^c$ where $n_{RC}^c$ is the number of RCs closed for a given arrangement. Since for all $t_{block}$ one has $\dot{n}_{RC,M}^c>\dot{n}_{RC,m}^c>0$, the enhancement of the absolute sensitivity must be traced back to the variation of the efficiency with respect to the number of closed RCs; indeed for $t_{block} < \bar{t} $ we have $d \eta_m/ d n_{RC}^c <d \eta_M/ d n_{RC}^c<0$. This relation shows that, at least in the first part of the transition region ($\bar{t} $ corresponds to $n_{RC}^c \sim 50 \%$) the maximum SPB configuration is not only the most efficient but it is also more resilient to RCs closure. This feature no longer holds for $t>\bar{t}$ and therefore when $n_{RC}^c \gtrsim 50 \%$ the sensitivity starts to decrease: $ 0 > d \eta_m/ dn_{RC}^c  > d \eta_M/ d n_{RC}^c $ and $\Delta \dot{ \eta} < 0$.

\begin{figure}[tb]
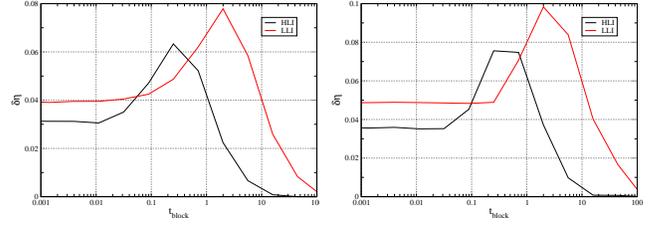

\includegraphics[width=0.48\columnwidth]{deltae_random_link_193_final.eps}
\includegraphics[width=0.48\columnwidth]{deltae_random_link_321_final.eps}
\caption{sensitivity for random networks (Left) $m=193$ (Right) $m=321$. A lower connectivity enhances the sensitivity}
\label{Fig.: sensitivity_random_link}
\end{figure}

As for the sensitivity of RHFs, we first notice that the initial sensitivities are very high: $21\%$ in
the HLI case and $37\%$ in the LLI case, while the enhancement of the sensitivity in the transition region is practically absent both in the HLI and LLI case.
This is due to a lower resilience to RCs closure in the case of the best arrangement (maximal DEG) and also has an impact on the overall performance of RHFs. Indeed a direct comparison shows that the most efficient configuration for RHF is much less resilient than for CT, i.e., $ |d\eta_M/dn_{RC}^c|_{RHF}  < |d\eta_M/dn_{RC}^c|_{CT}$ and this causes the RHF to be the network with the worst performance in the transition region (see also  fig.~\ref{Fig.: deltareal} where artificial networks' performances are compared with those of the real biological networks). \\
We finally  analyze the sensitivity of RNs.
We have that $\delta \eta$ for LLI varies between $5-10\%$ (see Fig.~\ref{Fig.: sensitivity_random_link}), and its behavior with $t_{block}$ is similar to the Cayley trees case.
The parameters that influence the sensitivity are both the stoichiometry and the average connectivity determined by the number $m$ of links eliminated from the original hexagonal lattice. Indeed, a smaller connectivity ($m=321$) makes the choice of the arrangement more crucial and therefore it enhances the sensitivity of the network, and this is true in particular in the transition region. \\
To sum up, the high values of the sensitivity by all networks analyzed indicate that the analysis of the topology of RCs arrangements becomes crucial in order to maximize the transport efficiency for such kind of networks, in particular for LLI networks. The behavior of the sensitivity with respect to $t_{block}$ is determined by the higher/lower resilience of efficient configurations to RC closure.

\subsection{Comparison with real networks and optimized arrangements}

In order to evaluate the optimality of the arrangements defined on the basis of centrality measures, we now compare the efficiency of all structures considered (CT, RHF and RNs) on one hand with those structures describing real, biological networks with a similar number of nodes and on the other hand with the results that one can obtain by searching for optimal arrangements by means of a numerical optimization  method (annealing).

\begin{figure}[htb]
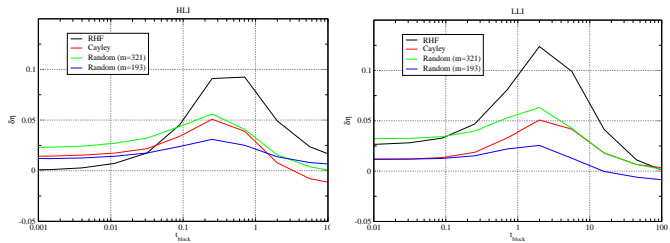

\begin{tabular}{cc}
\includegraphics[width=0.5\columnwidth]{deltareal_high.eps} &
\includegraphics[width=0.5\columnwidth]{deltareal_low.eps} \\
\end{tabular}
\caption{efficiency $\delta \eta_{real}$ vs. $t_{block}$ for different networks, HLI (right) LLI (left). The real biological networks have a comparable (active region) or better (transition region) performance than all other network topologies considered.}
\label{Fig.: deltareal}
\end{figure}

\begin{figure}[htb]
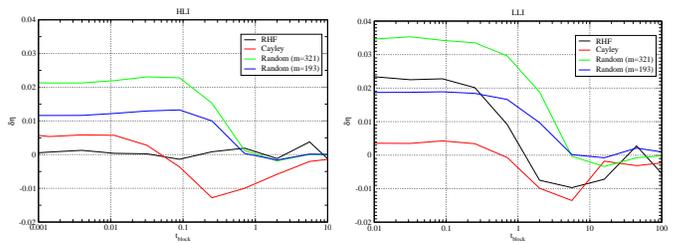

\begin{tabular}{cc}
\includegraphics[width=0.5\columnwidth]{deltaann_high.eps} &
\includegraphics[width=0.5\columnwidth]{deltaann_low.eps} \\
\end{tabular}
\caption{efficiency $\delta \eta_{annealing}$ vs. $t_{block}$ for different networks, HLI (right) LLI (left). For regular topologies the best LHI/RC arrangements based on centrality measures are optimal in that they allow to obtain approximately the same efficiencies obtained by numerical optimization; as for RNs the optimization procedure gives only slightly better results}
\label{Fig.: deltaannealing}
\end{figure}

To this aim in fig.~\ref{Fig.: deltareal} and~\ref{Fig.: deltaannealing} we plot $\delta_x=(\eta_x-\eta_M)/\eta_M$ where $x=$\{real, annealing\} and $\eta_M$ is the efficiency for the best arrangement of a given network.
A first relevant message of these plots is that the real biological networks are optimal both at LLI and HLI, in the sense that  they have an equal or better performance than all other network topologies considered, and the main differences arise in the first part of the transition region.
The optimality of real networks has to be related to their higher degree of connectivity with respect to the other networks analyzed. In particular they have a higher value of the average degree ($<k> \sim 5.75$ for both HLI and LLI), and smaller values of the average distance between pairs of nodes ($5.1$ for HLI and $4.7$ for LLI) and these features allow the excitons to explore greater parts of the network in search for an open RC; indeed, while for such networks $<d>_{cs}\in [5.5,7]$ is comparable or slightly higher than in the other cases , the exploration parameter reaches sensibly higher values: $<x_{cs}> \in [0.16-0.29]$ for HLI and $<x_{cs}> \in [0.17-0.28]$ for LLI, where the minimum corresponds to the active photosynthesis region and the maximum to the transition one.\\
Despite their lower connectivity, however, there are pairs of artificial network topologies and RCs arrangements that allow for efficiencies that are comparable with those of the real networks at least in the active region; this  in particular is true for Cayley trees for LLI conditions  with LH1/RC disposed in nodes of maximal SPB and RHF for HLI conditions and LH1/RC disposed in nodes of maximal degree.
As for the RNs,  those that better approximate the efficiency of the real ones are those with a higher degree of connectivity ($m=193$). \par
Our  analysis of artificial networks allows to identify the optimal pairs $\{\Gamma, \mathcal{A}\}$ that insure the realization of high efficient energy transport in all regions: (Cayley trees,max SPB),(RHF, max DEG), (random networks, max DEG). In order to further test the optimality of such arrangements based on centrality measures, we compare their efficiencies with those obtained through an optimization method. For the latter, we have used an optimization algorithm based on simulated annealing~\cite{Kirpkpatrick} to search for the best $\mathcal{A}$. The algorithm uses $1-\eta$ as a cost function, and evaluates it in the Master equation approach (see appendix B) in the limit of no RC closed ($t_{block}\ll R^{-1}$). Starting from a random arrangement, LH1 and LH2 positions are swapped till a minimum of the cost function is found.  The optimal arrangement $\mathcal{A}_{annealing}$ is then used for all values of $t_{block}$.
\\
The comparison (see fig.~\ref{Fig.: deltaannealing}) shows that, in particular for regular topologies (CT for HLI and LLI, RHF for HLI) the best LHI/RC arrangements based on centrality measures are optimal in that they allow to obtain, in all regions approximately the same efficiencies obtained by numerical optimization, while in the case of RNs the optimization procedure gives only slightly better results.

\section{Conclusions} \label{Sec : Conclusions}

In our work we have taken inspiration from biological light harvesting networks of purple bacteria and we have simulated an incoherent dissipative energy transport model devised for these systems on more general and abstract networks. Our analysis has focused on the subtle interplay between the global network structure and the arrangement of the two primary light harvesting complexes (LH1 and LH2) and their impact on the transport efficiency $\eta$. We have investigated well-known regular structures (Cayley trees and Regular Hyperbranched Fractals), as well as randomly-generated ones, and we have considered different network-theoretical centrality measures in order to select different LH1 and LH2 arrangements on the networks. \\
One can identify three transport regimes, depending on the relation between the rate of photon capture and the recycling time of the reaction centers (RC) which are placed in the center of LH1 complexes: the active, the transition and the saturated regime, characterized by the increasing average number of closed RCs found by an incoming exciton. \\
Our results clearly show that topology is crucial for efficient transport in the systems under analysis: different LH1/LH2 arrangements yield relative differences $\delta \eta$ in the transport efficiency which can be as high as $37\%$. In particular, the sensitivity of the efficiency to the LH1/LH2 arrangement $\delta \eta$ is enhanced when the number of LH1 is lower (low light illumination conditions) and typically in the transition region, where $50\%$ of the RCs is closed, making the LH1/LH2 arrangement even more relevant in this region. \\
Furthermore, we are able to identify guiding principles for optimal LH1/LH2 arrangement, given a global network topology. While efficient configurations always correspond to lower distances traveled by excitons to reach open RCs, the optimal criteria to achieve this feature strongly depend on
the topological properties of the networks considered. \\
For regular structures, efficient configurations can be effectively devised by means of 
centrality measures. Global (local) centrality measures induce efficient arrangements for networks with small (large) average distance between the nodes. Not only are these arrangements much more efficient than a random arrangement of complexes over the network: they are optimal, in the sense that they can match the efficiency obtained through a numerical optimization method. These results might be of use in the construction of artificial light harvesting networks where global control over the network structure and the arrangement of complexes is possible.\\
As for randomly-generated structures the arrangement which ensures the best efficiency is given by a local criterion (degree).
However, it is not possible to devise arrangements which are significantly more efficient than the random one, neither through centrality measures nor trough optimization.
This fact suggests that the transport process modeling purple bacteria membranes is robust with respect to limited amount of control in the structure design,
and it is optimized to take into account the disorder inherent in the construction of randomly-assembled structures.\\
Finally, real biological networks have in general equal or better performance with respect to the other structures considered.
While artificial structures with optimal arrangement of LH1/LH2 can ensure a comparable efficiency in the active region, a significant gap appears in the transition region where real biological networks substantially benefit from their higher level of connectivity. \\
\ \\
The framework we have developed in our work can be extended by comprising different centrality or analogous measures, and it can be applied to the study of more general processes of transport on complex networks with the presence of dissipation, trapping and congestion with the goal of deriving general design principles for artificial light harvesting networks. \par

\section*{Acknowledgements}

We thank A. Vespignani, C. Cattuto, A. Barrat, S. Fortunato, M. Caselle for useful discussions, comments, suggestions, and for their careful reading of the manuscript.

\appendix


\section{Characterization of the selected networks}

\begin{figure}[bt]
\centering
\begin{tabular}{cc}
\subfloat[DEG distribution CT]{\label{Fig.: Degree_distribution_Cayley}\includegraphics[width=0.2\textwidth]{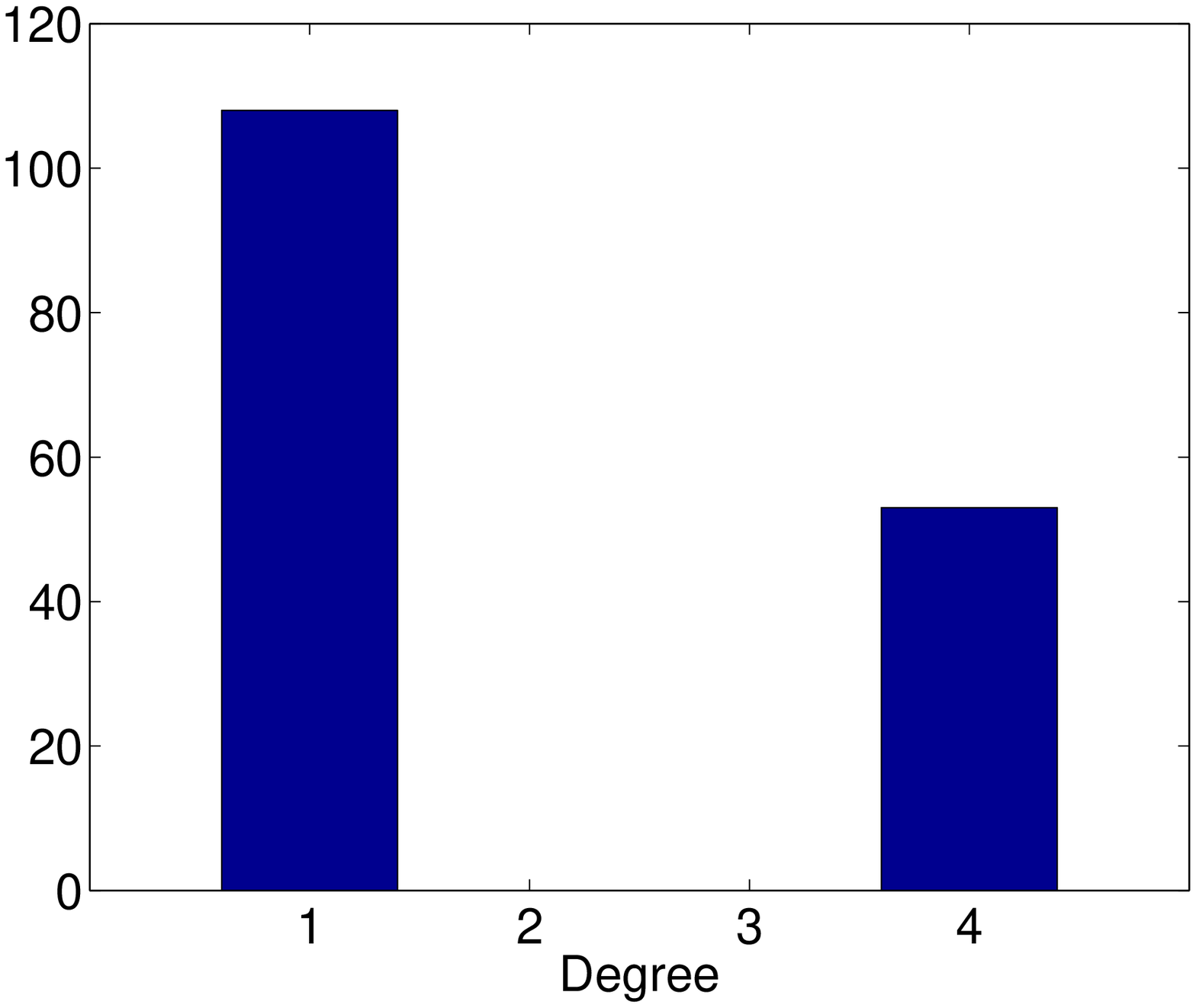}} &
\subfloat[SPB distribution
CT]{\label{Fig.: spb_distribution_Cayley}\includegraphics[width=0.2\textwidth]{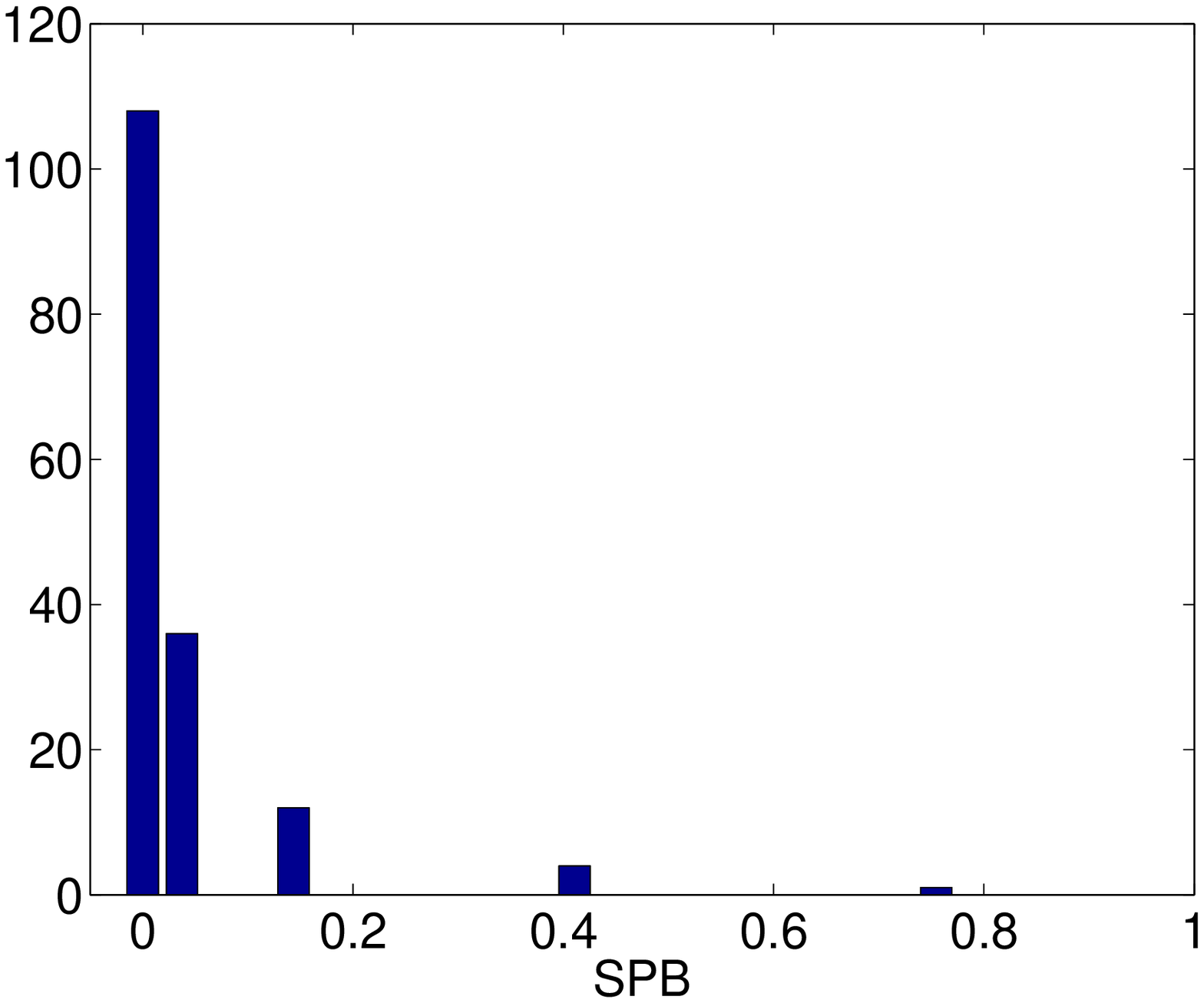}} \\
\subfloat[DEG distribution RHF]{\label{Fig.: Degree_distribution_RHF}\includegraphics[width=0.2\textwidth]{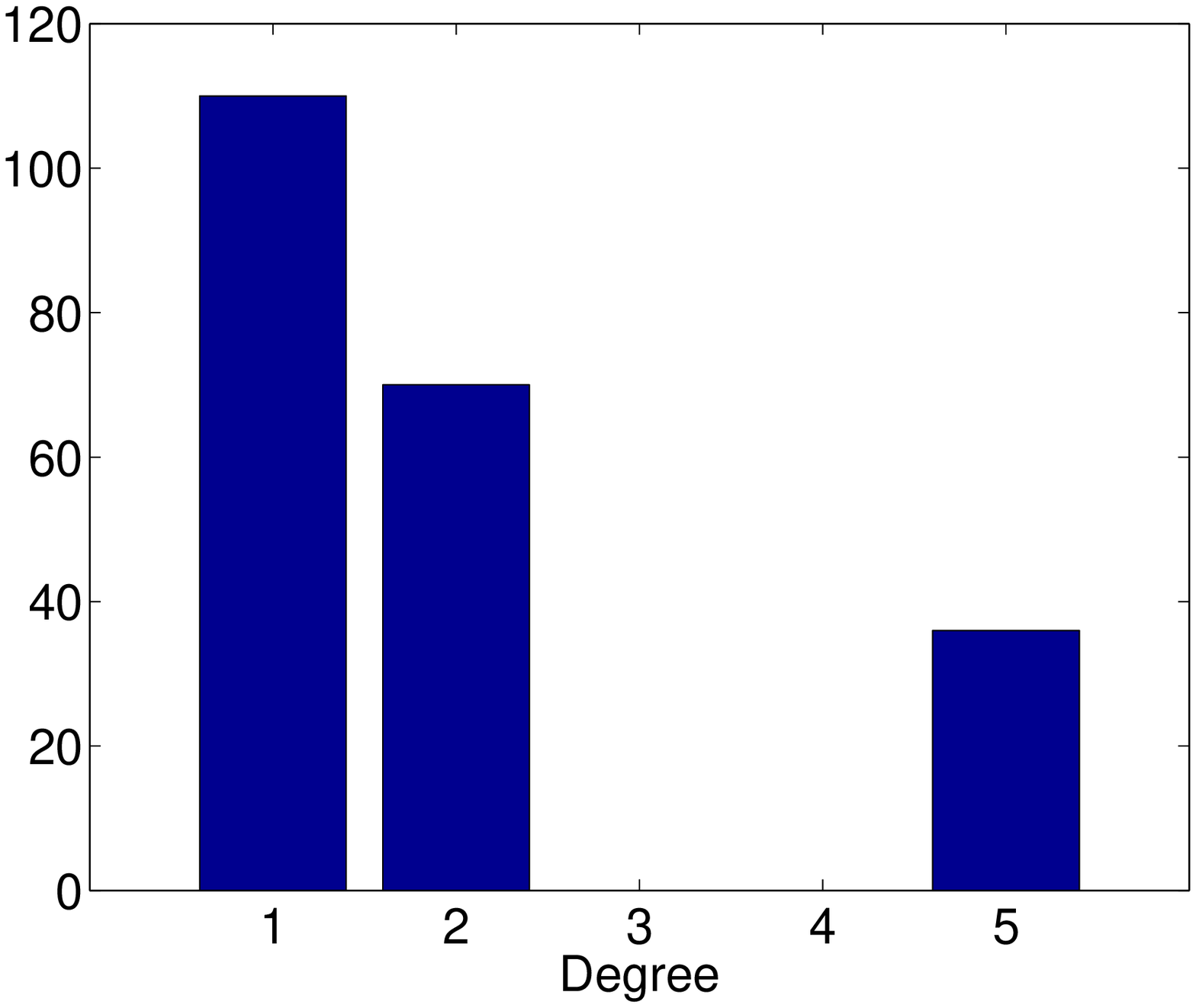}} &
\subfloat[SPB distribution
RHF]{\label{Fig.: spb_distribution_RHF}\includegraphics[width=0.2\textwidth]{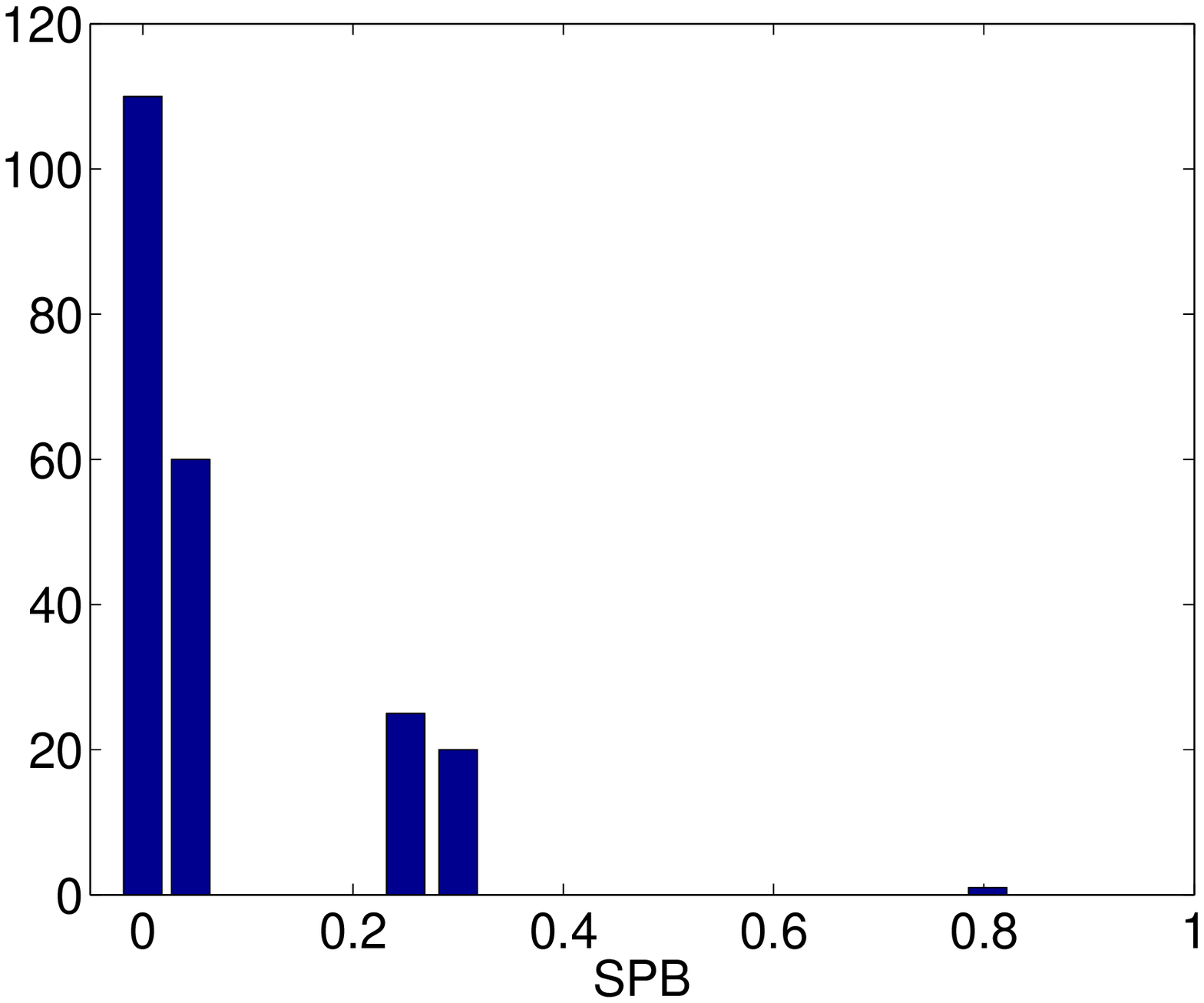}}
\end{tabular}
\caption{Distributions of DEG, SPB for CTs, $d=4$, $n=4$ and for RHF, $f=5$, $n=3$}
\label{Fig.: distributions regular}
\end{figure}

\begin{figure}[bt]
  \centering
\begin{tabular}{cc}
  \subfloat[DEG distribution real HLI]{\label{Fig.: Degree_distribution_real_HLI}\includegraphics[width=0.2\textwidth]{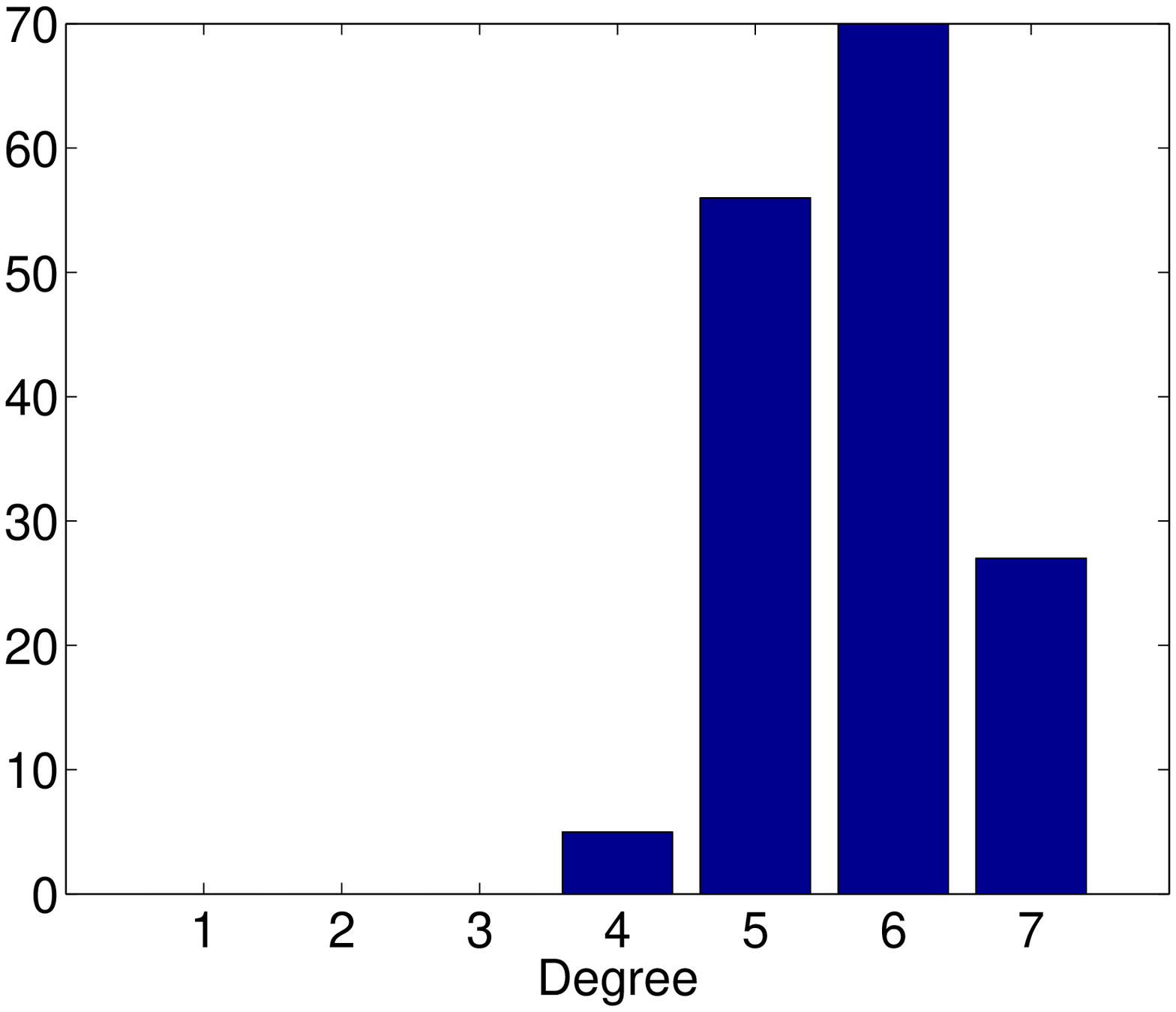}} &
  \subfloat[SPB distribution real  HLI]{\label{Fig.: spb_distribution_real_HLI}\includegraphics[width=0.2\textwidth]{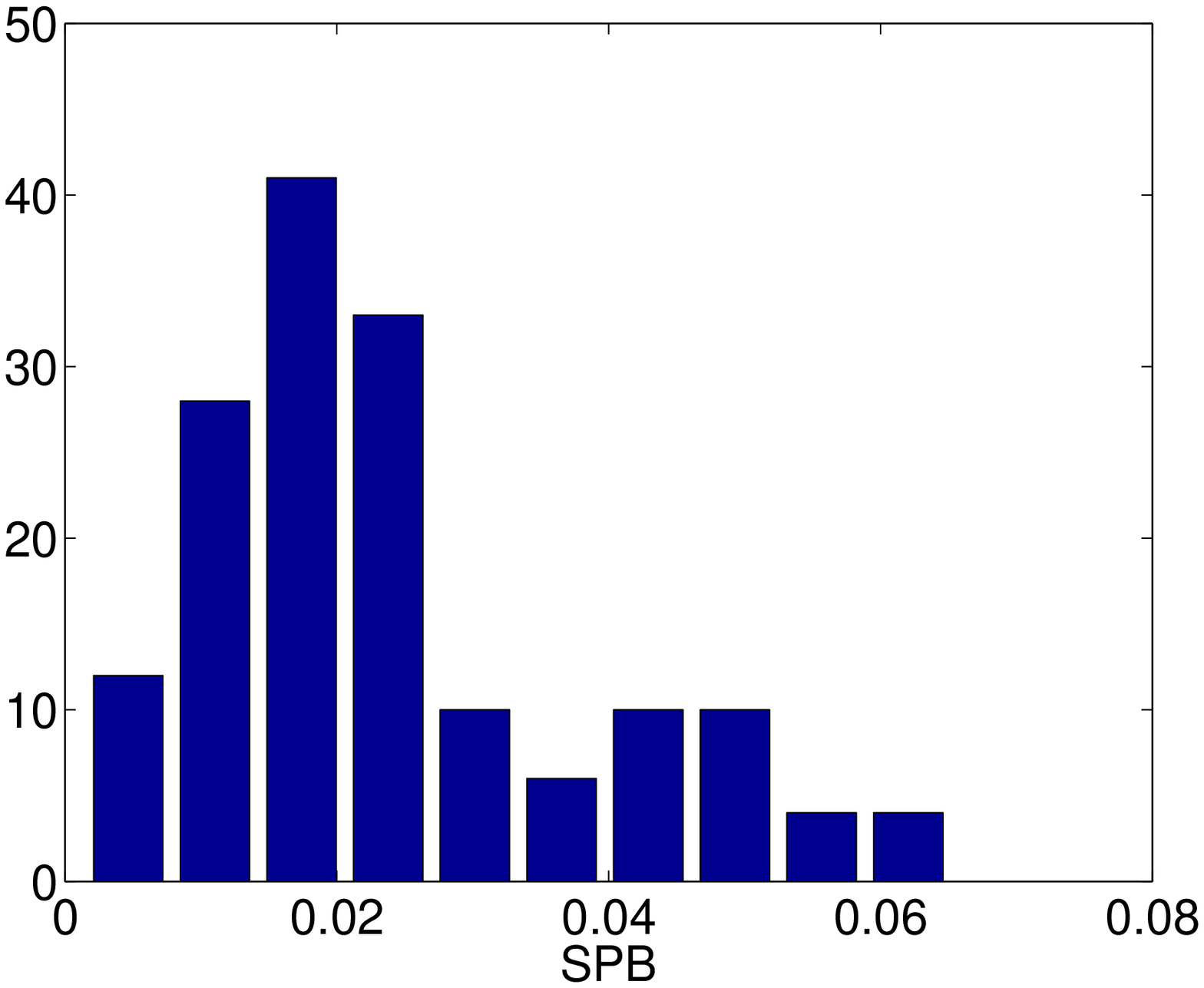}} \\
  \subfloat[DEG distribution real LLI]{\label{Fig.: Degree_distribution_real_LLI}\includegraphics[width=0.2\textwidth]{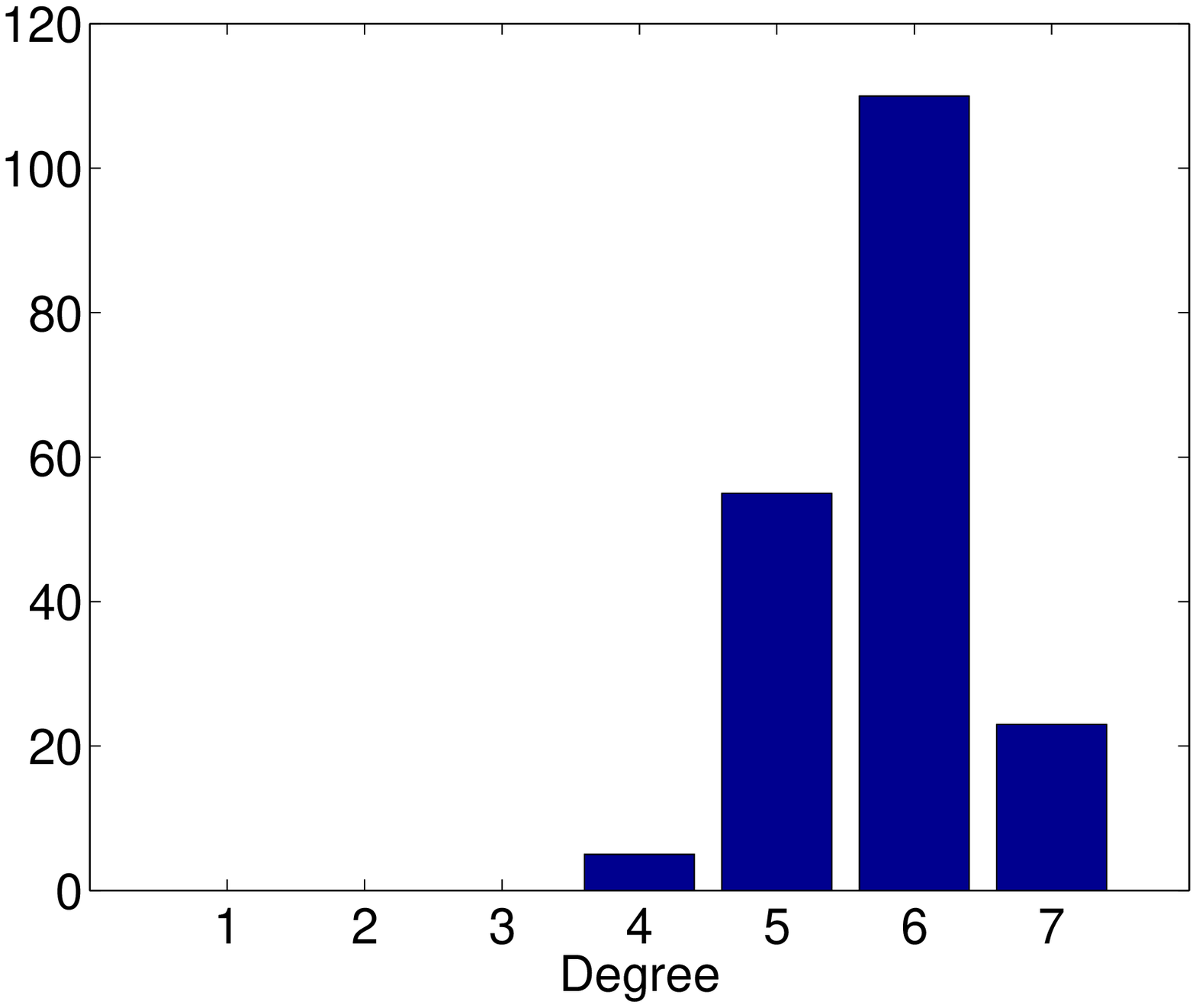}} &
  \subfloat[SPB distribution real  LLI]{\label{Fig.: spb_distribution_real_LLI}\includegraphics[width=0.2\textwidth]{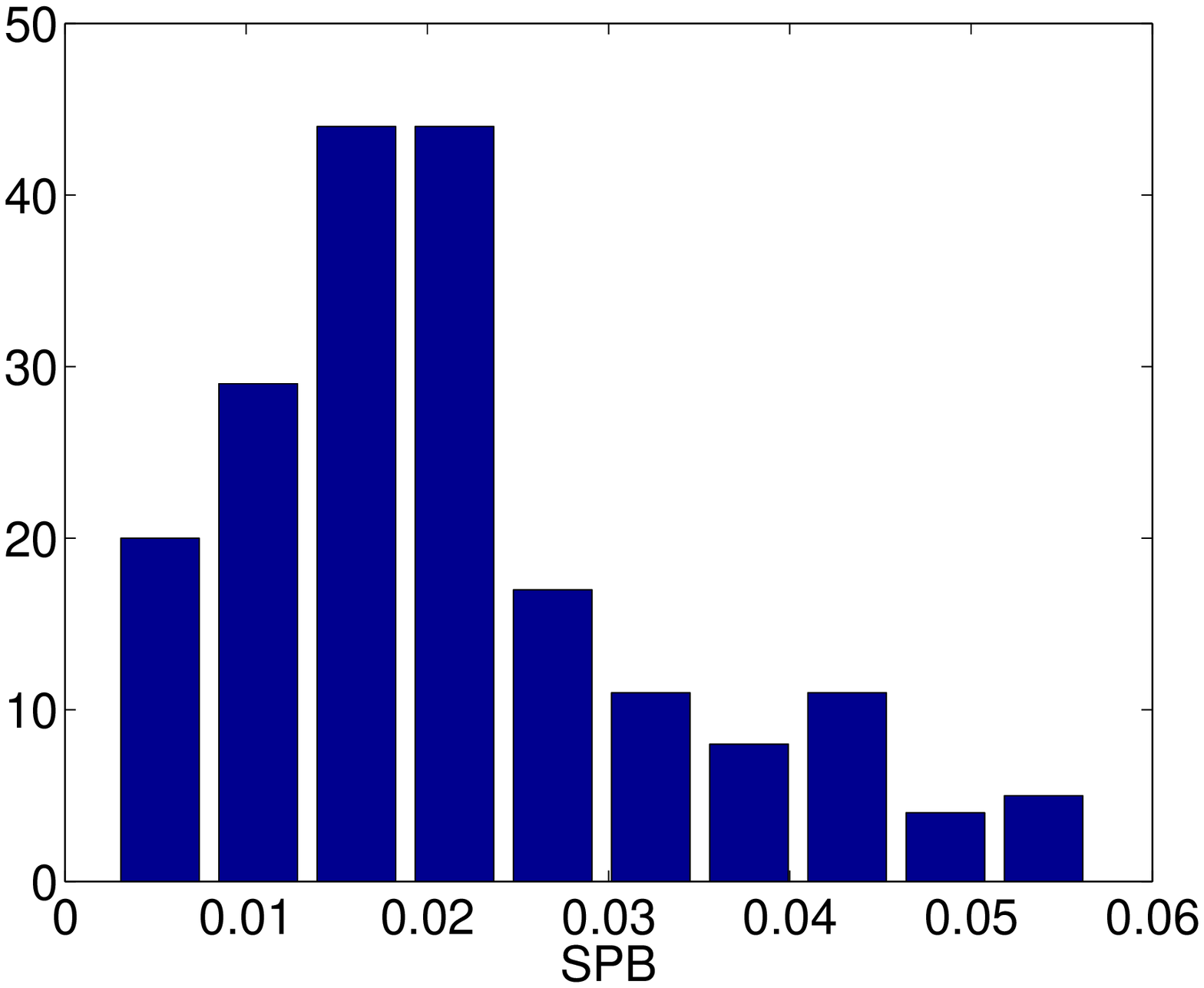}}
\end{tabular}
 \caption{Distributions of DEG, SPB for real HLI and LLI networks}
  \label{Fig.: animals-HLI}
\end{figure}

\begin{figure}[bt]
\centering
\begin{tabular}{cc}
\subfloat[DEG distribution $m=193$]{\includegraphics[width=0.2\textwidth]{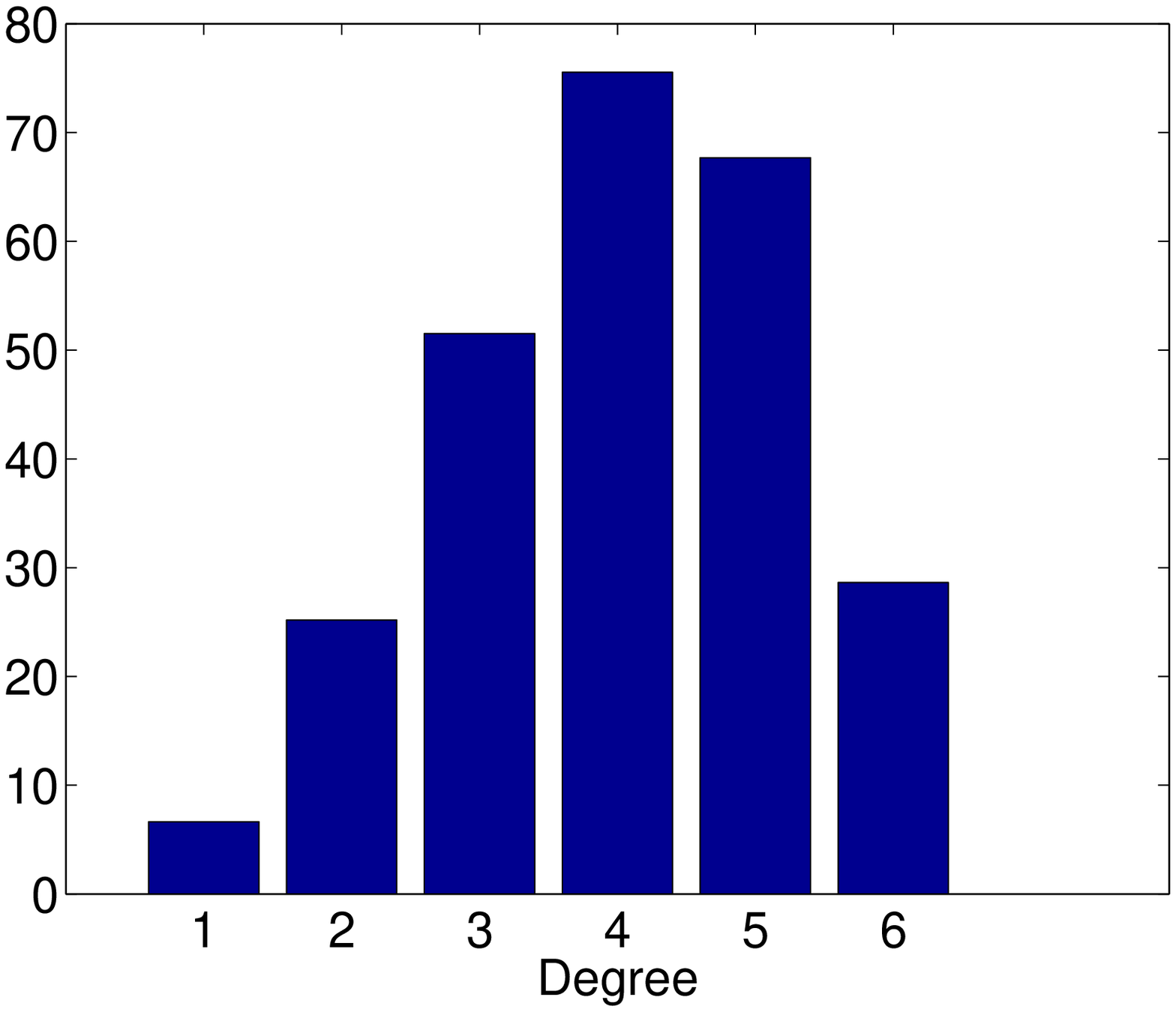}} &
\subfloat[SPB distribution $m=193$]{\includegraphics[width=0.2\textwidth]{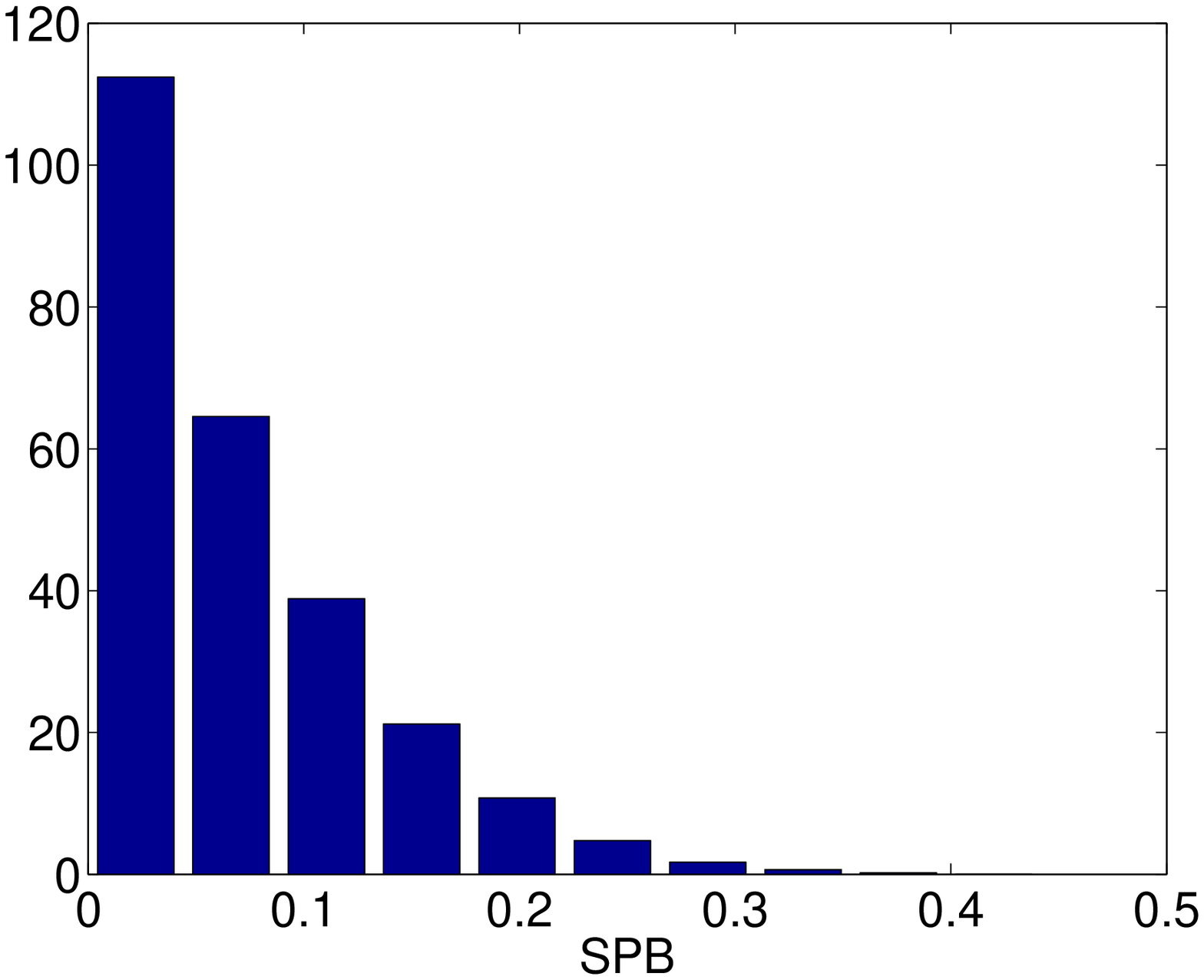}} \\
\subfloat[DEG distribution $m=321$]{\includegraphics[width=0.2\textwidth]{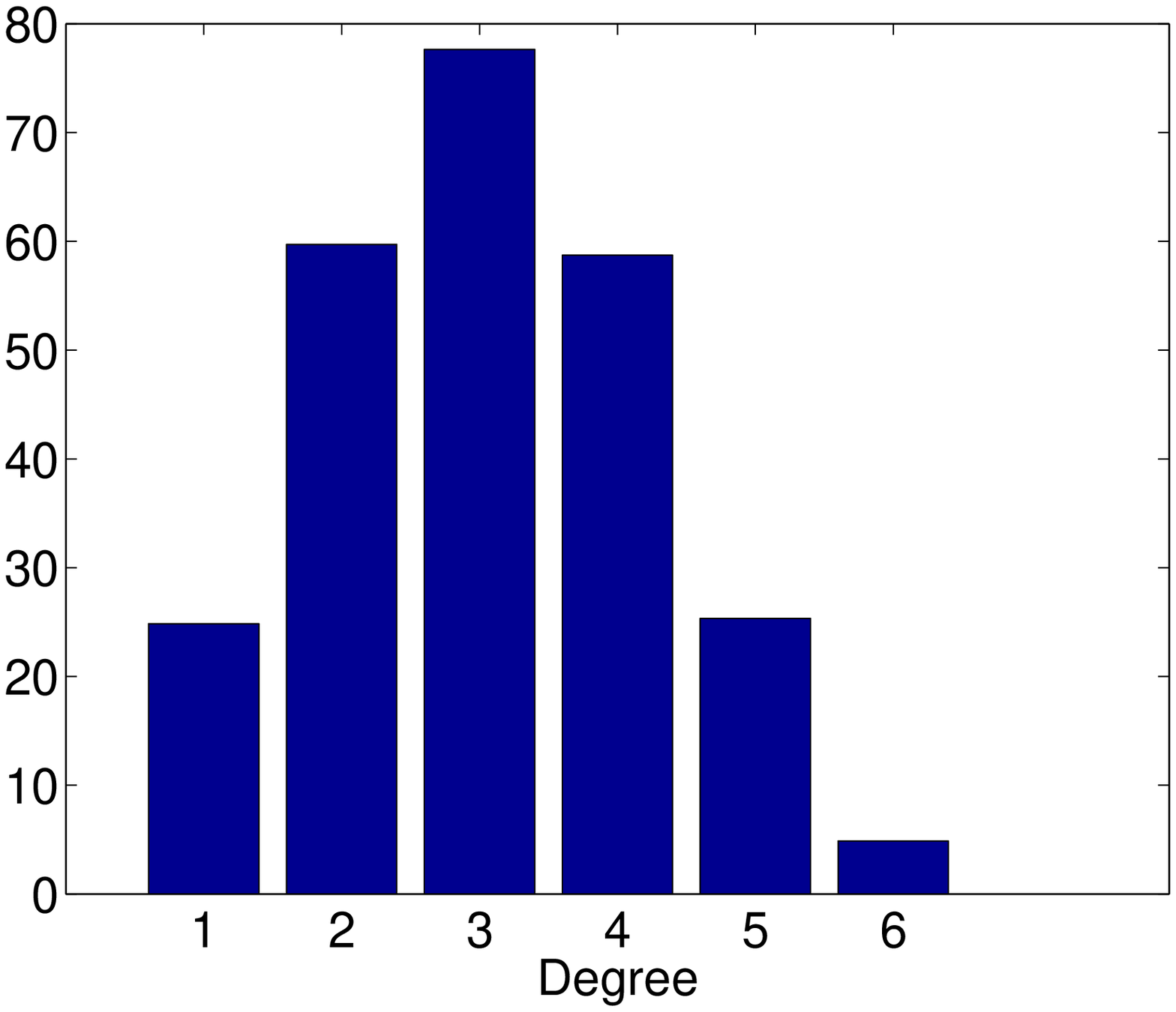}} &
\subfloat[SPB distribution $m=321$]{\includegraphics[width=0.2\textwidth]{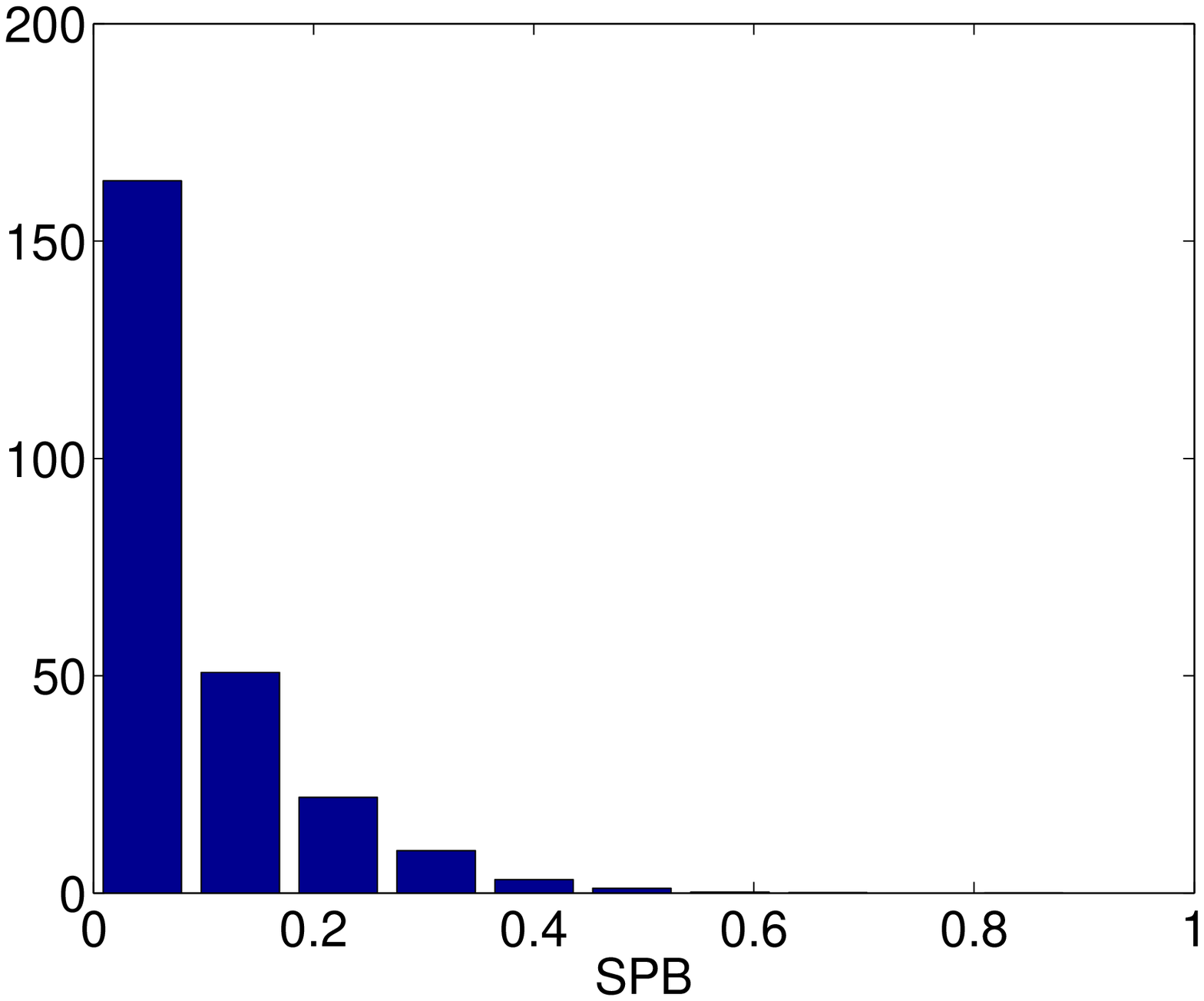}}
\end{tabular}
\caption{Distributions of DEG, SPB for random networks}
\label{Fig.: distributions random}
\end{figure}

We report in the figures below (Fig.~\ref{Fig.: distributions regular},\ref{Fig.: animals-HLI},\ref{Fig.: distributions random}) the distribution of the network-theoretical measures (DEG, SPB) for the selected networks. Notice that the distributions shown for random RNs (Fig.~\ref{Fig.: distributions random}) are an average over
$M=100$ networks with different topology.

\ \\
As for the CTs: the DEG distribution is trivial, there are $d(d-1)^{n-1}=108$ nodes with DEG$=1$ (peripheral nodes) while all the others have DEG$=d=4$; SPB is a strongly decreasing function of the generation of the nodes (their distance from the center). The number of nodes of generation $k$ scales as $\sim d^k$, so most nodes (144) have a small SPB ($< 0.1$), while 12 nodes have SPB $\sim 0.2$, 4 nodes have SPB $\sim 0.4$, and one single node (the central one) has SPB $\sim 0.8$). The DEG and the SPB are poorly correlated (linear correlation coefficient  $c=0.57$).\\
As for the RHF: the DEG distribution is simple, a number $(f+1)^{n-1}=36$ of the nodes (the central nodes of the component star graphs) have DEG$=f=5$, a number $\frac{(f^{n+1}-1)}{(f-1)(f+1)^n}=26$ have DEG$=2$ and the remaining ones have DEG$=1$; SPB is again a decreasing function of the distance from the central node, the majority of nodes (166) have SPB $< 0.1$ while a smaller fraction has values SPB in the range $0.2-0.4$, and one node (the central one) has a SPB $\sim 0.8$). Again, the DEG and the SPB are poorly correlated ($c=0.47$). \\
As for the RNs: the DEG distribution is peaked around DEG$\sim 4$, DEG$\sim 3$, corresponding to $m=197$, $m=321$ and it is skew, showing a tail for small/high DEG $m=197$, $m=321$; the SPB distribution is decreasing, all nodes have small values of SPB (quite differently than in the case of CT and RHF), values of SPB ($\sim 0.2-0.3$) are attained only for a few nodes. The DEG and the SPB are poorly correlated  ($c=0.60$ for $m=197$ and
$c=0.55$ for $m=321$). \\
As for the real biological networks: the DEG distribution is sharply peaked around DEG$\sim 6$; the SPB distribution is irregular, all nodes have small values of SPB ($0.01-0.06$), with most node having SPB $\sim 0.02$.
The DEG and the SPB display a weak correlation for LLI networks ($c=0.70$) and a stronger one for HLI networks ($c=0.82$).

\section{Master equation approach\label{MEapp}}

Alternatively to performing random walk simuations, one can evaluate many figures of merit directly from a Master equation approach.
The single exciton transfer and trapping can be described in terms of the random migration of a single exciton in a membrane with an effective fraction of closed RC. Indeed, for any fixed $t_{block}$ the number of closed RC rapidly reaches its average value, so that
each exciton sees an effective fraction of closed RCs corresponding to a this average value~\cite{Fassioli}.
This allows for a Master equation treatment of both active (all RCs open) and saturated (most RCs closed) photosynthesis.
More precisely, the process is governed by different probability rates and thus can be modeled by a (Markovian) Master Equation (ME)
\be
\frac{d p_m }{d t} = \sum_n K_{mn} p_n
\ee
whose solution is $\ket{p(t)} = e^{Kt} \ket{p(0)}$. The $N$-dimensional vector $\ket{p}$ is composed by the probabilities $\{p_n\}$ that the excitons be in site $n$. The initial probabilities are given by
\be
\ket{p(0)}= {p_n (0)} ,
\qquad p_n(0) = \left\{ \begin{array}{c}
\sigma_1/\sigma \quad \mbox{if } n = LH1 \\
\sigma_2/\sigma \quad \mbox{if } n = LH2 \\
0 \ \quad \mbox{if } n = RC
\end{array} \right.
\ee
and transfer matrix $K$ can be written as \cite{Fassioli}:
\bae
K_{mn}&=&W_{mn}-\delta_{mn}(\sum_l W_{ln} + \delta_{n,RC^o} k_{cs})-\\ \nonumber
&=& - \delta_{mn}(k_{diss}(1-\delta_{n,RC^c})+k^*_{diss}\delta_{n,RC^c})
\eae
where: The exciton transfer rates $W_{mn}$ are taken as the inverse of the exciton transfer times given in Table \ref{Tab.: exciton transfer times}, $k_{diss}=t_{diss}^{-1}$, $k^*_{diss}=t_{RC^c_{diss}}^{-1}$, $k_{cs}=t_{cs}^{-1}$.
In the ME approach,  the effect of the recycling time of RCs can be taken into account by keeping $N_{block}$ RCs closed, where $N_{block}$ is determined as the average number of RCs closed at time $t$ in a fixed illumination condition. For each value of $N_{block}$
some relevant functionals (efficiency, exciton lifetime, etc.: See below) are evaluated. The procedure is repeated a high number of times, randomly choosing the RC that are closed, and one finally determines the average of the functionals for a given $N_{block}$
\footnote{The exact relation between $N_{block}$ and $t_{block}$ is obtained by $t_{block}$ via $t_{block} = \frac{h \nu \cdot N_{block}}{\eta(N_{block}) \cdot I \cdot \sigma}$ where $\eta(N_{block})$ is the quantum yield of the process (see below) when $N_{block}$ RC are closed}.\\
In the ME approach, $\eta$ can be simply obtained from the inverse transfer matrix $K^{-1}$~\cite{Fassioli}. Indeed we have $\eta = \int_0^\infty \omega_{cs}(t) dt $ where $\omega_{cs}(t)$ is the probability of an exciton causing charge separation at time $t$ and is given by $ \omega_{cs} = k_{cs} \sum_{n \in RC} p_n (t) =  k_{cs} \sum_{n \in RC} \bra{n} e^{Kt} \ket{p(0)} $ where $\ket{n}$ is the probability vector corresponding to an exciton being in site $n$ with certainty. Therefore $\eta = -k_{cs} \sum_{n \in RC } \bra{n}  K^{-1} \ket{p(0)} $ . \\
The results of the random walk dynamics have been compared with the ones obtained in the Master equation approach and the two methods show complete agreement.


\section*{References}
\begin{thebibliography}{100}
\bibitem{Photo} D. O. Hall, K. K. Rao, \textit{Photosynthesis}, Cambridge University Press, 2009;
R. E. Blankenship, \textit{Molecular mechanisms of photosynthesis}, John Wiley \& Sons, 2002.

\bibitem{Structureofmembranes} S. Scheuring \textit{et al.}, Proc. Natl. Acad. Sci. USA \textbf{101}, 11293 (2004); Bahatyrova \textit{et al.}, , Nature \textbf{430}, 1058 (2004); S. Scheuring, and J. N. Sturgis, Photosynth. Res. \textbf{102}, 197 (2009); X. Hu \textit{et al.}, Proc. Natl. Acad. Sci. USA. \textbf{95}, 5935 (1998).
\bibitem{LHCIIstructure} Z. Liu \textit{et al.}, Nature {\bf 428}, 287 (2004);
Doust, A. B. \textit{et al.}, J. Mol. Biol. \textbf{344}, 135 (2004).




\bibitem{transfertimes}
R. van Grondelle, J. Dekker, T. Gillbro, and V. Sundstrom, Biochim. Biophys. Acta 1187 (1994); S. Hess, M. Chachisvilis, K. Timpmann, M. Jones, G. Fowler, C. Hunter, and V. Sundstrom, Proc. Natl. Acad. Sci. USA \textbf{92}, 12333 (1995); X. Hu, T. Ritz, A. Damjanovic, F. Autenrieth and K. Schulten, Quart. Rev. of Biophysics \textbf{35}, 1 (2002); A. Damjanovic, T. Ritz, and K. Schulten, Int. J. Quantum Chem. \textbf{77}, 139 (2000);

\bibitem{Ritz} T. Ritz, S. Park and K. Schulten, J. Phys. Chem. B \textbf{105}, 8259 (2001).

\bibitem{archi} In Ref. \cite{Ritz} Ritz \emph{et al.} consider two basic architectures, which can be considered as extreme cases of possible arrangements of the  photosynthetic unit: the \textit{stripe} architecture where LH1-RC complexes form a 2-fold array and the \textit{circular} architecture where an LH1-RC complex is surrounded by 9 LH2's and is isolated from other LH1-RC complexes. The architectures investigated by Johnson \emph{et al.} are very similar.


\bibitem{Sturgis1} S. Sheuring, J.-L. Rigaud and J. N. Sturgis, EMBO J. \textbf{23}, 4127 (2004); S. Scheuring and J. N. Sturgis, Science \textbf{309}, 484 (2005);

\bibitem{Schulten} K. Schulten, in \textit{Simplicity and complexity in proteins and nucleic acids}, Dahlem University Press (1999).


\bibitem{Fassioli} F. Fassioli, A. Olaya-Castro, S. Scheuring, J. N. Sturgis and N. F. Johnson, Bioph. J. \textbf{97}, 2464 (2009).

\bibitem{Johnson} F. Caycedo-Soler, F. J. Rodriguez, L. Quiroga and N. F. Johnson, N. J. Phys. \textbf{12}, 095008 (2010).

\bibitem{QBioExperiments}
G. S. Engel \textit{et al.}, Nature \textbf{446}, 782 (2007); I. P. Mercer \textit{et al.}, Phys. Rev. Lett. \textbf{102}, 057402 (2009); E. Collini \textit{et al.}, Nature \textbf{463}, 644 (2010);

\bibitem{QBioTheory}
M. Mohseni \textit{et al.}, J. Chem. Phys. {\bf 129}, 174106 (2008);  \\
A. Olaya-Castro \textit{et al.}, Phys. Rev. B \textbf{78}, 085115 (2008); \\
F. Caruso et al., J. Chem. Phys. \textbf{131}, 105106 (2009); \\
A. Ishizaki, G.R. Fleming, Proc. Natl. Acad. Sci. USA \textbf{106}, 17255 (2009); \\
P. Giorda, S. Garnerone, P. Zanardi, and S. Lloyd, arXiv:1106.1986 (2011).
\bibitem{OlayaScholesETreview} A. Olaya-Castro, and G. Scholes, Int. Rev. Phys. Chem. \textbf{30}, 49 (2011).
\bibitem{Escalante} N. Reynolds \textit{et al.}, J. Am. Chem. Soc. \textbf{129}, 14625 (2007);
M. Escalante \textit{et al.}, J. Am. Chem. Soc. \textbf{130}, 8892 (2008);  M. Escalante \textit{et al.}, Nano Lett. \textbf{10}, 1450 (2010);

\bibitem{DendrimersReviews} P. E. Froehling, Dyes and Pigments \textbf{48} 187 (2001); D. S. Bradshaw, and D. L. Andrews, Polymers \textbf{3}, 2053 (2011).

\bibitem{notealgorithm} The detailed description of the exact simulation algorithm can be found in \cite{Fassioli}.

\bibitem{Blumen1} G.H. Kohler and A. Blumen, J. Phys. A \textbf{23} (1990).
\bibitem{vari}
S. Raychaudhuri, Y. Shapir, V. Chernyak, and S. Mukamel, Phys. Rev. Lett. \textbf{82} (2000).

\bibitem{Blumen2} A. Blumen, A. Jurjiu, Th. Koslowski, and Ch. von Ferber, Phys. Rev. E \textbf{67}, 061103 (2003);
 F. Jasch, Ch. von Ferber, and A. Blumen, Phys. Rev. E \textbf{68}, 051106 (2003).

\bibitem{Muelken} O. M\"ulken, V. Bierbaum, and A. Blumen, J. Chem. Phys. \textbf{124}, 124905 (2006); A. Blumen, V. Bierbaum, and O. Mülken, Physica A \textbf{371}\textbf{}, 10 (2006); O. M\"ulken, and A. Blumen, Physica E \textbf{42}, 576 (2010); O. M\"ulken, and A. Blumen, Phys. Rep. \textbf{502}, 37 (2011).

\bibitem{complex} M. E. J. Newman, \textit{Networks: an introduction}, Oxford University Press (2010); A. Barrat, M. Barth\'elemy, A. Vespignani, \textit{Dynamical processes on complex networks}, Cambridge University Press (2010).

\bibitem{Borgatti} S. P. Borgatti, M. G. Everett, Social networks \textbf{28}, 466 (2006).

\bibitem{Freeman1} L. C. Freeman, Social networks \textbf{1}, 215 (1979).

\bibitem{Freeman2} L. C. Freeman, Quality and quantity \textbf{14}, 585 (1980).

\bibitem{Newman} M. E. J. Newman, arXiv:0309045v1 (2003).

\bibitem{Brandes} U. Brandes, J. Math. Sociology \textbf{25}, 163 (2001).

\bibitem{Thadakamalla} H. P. Thadakamalla, R. Albert, and S.R.T. Kumara, Phys. Rev. E \textbf{72}, 066128 (2005).

\bibitem{NoteSPB(d)1} Notice that the measure introduced in ~\cite{Thadakamalla} was restricted to $d=2$ and was evaluated for the first neighbors of the node, instead of the node itself.

\bibitem{NoteSPB(d)2} We notice that the values of $SPB(d)$ become more and more correlated with the degree for decreasing $d$, and thus give progressively better arrangements since the max DEG arrangement is optimal for RHF.

\bibitem{Kirpkpatrick} S. Kirkpatrick, C. D. Gelatt, Jr. and M. P. Vecchi, Science \textbf{220}, 671 (1983);
P. van Laarhoven, E. Aarts, \textit{Simulated annealing: theory and applications}, Kluwer (1988).


\end{thebibliography}
\end{document}